\input{amssymb.sty}
\documentstyle[aps,pre,floats,epsf,eqsecnum,epsfig]{revtex}

\def\Tc{T_{\mbox{\scriptsize c}}}
\def\rhoc{\rho_{\mbox{\scriptsize c}}}
\input{psfig.sty}
\begin{document}
\draft

\twocolumn[\hsize\textwidth\columnwidth\hsize\csname@twocolumnfalse\endcsname 

\title{Asymmetric Fluid Criticality I: Scaling with Pressure Mixing}

\author{Young C. Kim, Michael E. Fisher and G. Orkoulas\footnote{Now at: Department of Chemical Engineering, Rensselaer Polytechnic Institute, Troy, New York 12180}}
\address{Institute for Physical Science and Technology, \\ University of Maryland, College Park, Maryland 20742}

\date{\today}

\maketitle
\begin{abstract}

The thermodynamic behavior of a fluid near a vapor-liquid and, hence, asymmetric critical point is discussed within a general ``complete'' scaling theory incorporating pressure mixing in the nonlinear scaling fields as well as corrections to scaling. This theory allows for a Yang-Yang anomaly in which $\mu_{\sigma}^{\prime\prime}(T)$, the second temperature derivative of the chemical potential along the phase boundary, diverges like the specific heat when $T\rightarrow T_{\mbox{\scriptsize c}}$; it also generates a leading singular term, $|t|^{2\beta}$, in the coexistence curve diameter, where $t\equiv (T-T_{\mbox{\scriptsize c}})/T_{\mbox{\scriptsize c}}$. The behavior of various special loci, such as the critical isochore, the critical isotherm, the $k$-inflection loci, on which $\chi^{(k)}\equiv \chi(\rho,T)/\rho^{k}$ (with $\chi = \rho^{2}k_{\mbox{\scriptsize B}}TK_{T}$) and $C_{V}^{(k)}\equiv C_{V}(\rho,T)/\rho^{k}$ are maximal at fixed $T$, is carefully elucidated. These results are useful for analyzing simulations and experiments, since particular, nonuniversal values of $k$ specify loci that approach the critical density most rapidly and reflect the pressure-mixing coefficient. Concrete illustrations are presented for the hard-core square-well fluid and for the restricted primitive model electrolyte. For comparison, a discussion of the classical (or Landau) theory is presented briefly and various interesting loci are determined explicitly and illustrated quantitatively for a van der Waals fluid.
\end{abstract}
\vspace{.3in}
]

\section{Introduction and Overview}
\label{sec1}

In 1964 Yang and Yang \cite{yang:yang} derived the thermodynamic relation
 \begin{equation}
  C_{V}^{\mbox{\scriptsize tot}}(T,V) = VT\left(\frac{\partial^{2} p}{\partial T^{2}}\right) - NT\left( \frac{\partial^{2} \mu}{\partial T^{2}} \right), \label{yang-yang}
 \end{equation}
for a fluid at pressure $p$, chemical potential $\mu$, and temperature $T$, where $V$ is the volume and $N$ the number of particles, while $C_{V}^{\mbox{\scriptsize tot}}$ is the constant-volume specific heat or, better, heat capacity. This has since been called the Yang-Yang relation \cite{fis:ork}. When it is applied to the two-phase region beneath the critical temperature $T_{\mbox{\scriptsize c}}$, one has $p = p_{\sigma}(T)$ and $\mu = \mu_{\sigma}(T)$ (where $\sigma$ denotes the phase boundary on which liquid and vapor may coexist), and the partial derivatives become total derivatives. Since the observations of Voronel' and coworkers in 1962-63 \cite{voronel} it has been well established that the heat capacity $C_{V}^{\mbox{\scriptsize tot}}(T)$ diverges weakly along the phase boundary when the critical point is approached. The divergence of $C_{V}^{\mbox{\scriptsize tot}}$ then implies that one, the other, {\em or} both of the second derivatives $p_{\sigma}^{\prime\prime}(T)$ and $\mu_{\sigma}^{\prime\prime}(T)$ must diverge when $T \rightarrow T_{\mbox{\scriptsize c}}-$ along the phase boundary. The lattice gas model and its standard variants predict that $\mu_{\sigma}^{\prime\prime}$ remains finite at $T_{\mbox{\scriptsize c}}$ while $p_{\sigma}^{\prime\prime}$ diverges like the specific heat \cite{lee:yan}. However, Yang and Yang suggested that in real fluids {\em both} should diverge \cite{yang:yang}: clearly this is a basic issue for the description and understanding criticality in fluids.

Recently, Fisher and coworkers \cite{fis:ork} carefully analyzed experimental two-phase heat-capacity data for propane (C$_{3}$H$_{8}$) and CO$_{2}$ and showed that the evidence rather strongly indicates that $\mu_{\sigma}^{\prime\prime}$ does {\em indeed} diverge like the specific heat when $T\rightarrow \Tc -$. They dubbed this phenomenon a {\em Yang-Yang anomaly} \cite{fis:ork}. Even though impurities in the propane system have definite effects on the heat-capacity data \cite{kos:ani}, the existence of a Yang-Yang anomaly cannot be ruled out and , in our assessment, remains the most plausible scenario. In fact, Orkoulas {\em et al.} \cite{ork:fis:pan} have performed grand canonical Monte Carlo simulations for the hard-core square-well fluid and concluded that this model system probably exhibits a negative but small Yang-Yang anomaly, {\em i.e.}, a specific-heat-like divergence in the chemical potential derivative, $d^{2}\mu_{\sigma}/dT^{2}$, of magnitude significantly less than the divergence of $v(d^{2}p_{\sigma}/dT^{2})$, where $v=V/N$. How can one then accommodate such a Yang-Yang anomaly in scaling theory?

The concept of asymptotic scaling has proved a powerful tool for gaining insight into critical phenomena in a variety of systems including fluids \cite{widom,fisher3,kadanoff,griffiths,patashinskii,widom2}. Furthermore, a scaling equation of state has been rather well confirmed experimentally for many fluids \cite{sengers,sengers2}. The currently accepted asymptotic scaling description of fluid criticality \cite{mer} requires two scaling fields, namely a thermal field, $\tilde{t}$, and an ordering field, $\tilde{h}$, that, in leading order, are both linear combinations of $~t\propto T-\Tc~$ and $~h = \mu - \mu_{\mbox{\scriptsize c}},~$ the deviations of the temperature and chemical potential from their critical values. This description has also been extended to describe fluctuations in finite systems and applied to estimating the critical points of model fluids \cite{bruce:wilding}. However, within this scaling description, the chemical potential is always analytic along the phase boundary and through the critical point, thereby having a {\em finite} value of the second temperature derivative, $\mu_{\sigma}^{\prime\prime}$. Hence, in order to account for a Yang-Yang anomaly, the current scaling description must be modified. Indeed, Fisher and Orkoulas \cite{fis:ork} argued that the pressure deviation, $~p-p_{\mbox{\scriptsize c}},~$ must mix into the scaling fields, especially into the ordering field $\tilde{h}$.

In order to formulate a scaling theory in which the pressure is mixed into the scaling fields, let us consider a {\em full thermodynamic description} of a one-component fluid as provided by a functional relation between the three thermodynamic fields, pressure, $p$, chemical potential, $\mu$, and temperature, $T$, say
 \begin{equation}
  \Upsilon(p,\mu,T) = 0.  \label{thermo}
 \end{equation}
Extending the approach sketched in Ref.\ [2] to include the full spectrum of correction exponents (see also \cite{kim:fis:bar,fis:bar}) one may formulate a rather general scaling hypothesis by asserting that near a typical critical point, $(p_{\mbox{\scriptsize c}},\mu_{\mbox{\scriptsize c}},\Tc)$, the thermodynamics can be described, at least asymptotically, by
 \begin{equation}
  \Psi(\lambda^{2-\alpha}\tilde{p},\: \lambda \tilde{t},\:\lambda^{\Delta}\tilde{h};\:\lambda^{-\theta_{4}} u_{4},\:\lambda^{-\theta_{5}}u_{5},\cdots) = 0, \label{scaling-descript}
 \end{equation}
where $\lambda$ is a free, positive scaling parameter. In this expression $\tilde{p}(p,\mu,T)$, $\tilde{t}(p,\mu,T)$ and $\tilde{h}(p,\mu,T)$ are the three relevant {\em nonlinear scaling fields}, while $u_{4}(p,\mu,T)$ and $u_{5}(p,\mu,T)$ are the leading even and odd irrelevant scaling fields, respectively. Including the quadratic nonlinear terms, we may write the basic nonlinear fields as
 \begin{eqnarray}
  \tilde{p} & = & \check{p} - k_{0}t - l_{0}\check{\mu} \nonumber \\
   &  & -\: r_{0}t^{2} - q_{0}\check{\mu}^{2} - v_{0}t\check{\mu} - m_{0}\check{p}^{2} - n_{0}\check{p}t - n_{3}\check{p}\check{\mu} \nonumber \\
   &  & +\: O_{3}(t,\check{\mu},\check{p}),  \label{tilde-g}  \\
  \tilde{t} & = & t - l_{1}\check{\mu} - j_{1}\check{p} \nonumber \\
   &  & -\: r_{1}t^{2} - q_{1}\check{\mu}^{2} - v_{1}t\check{\mu} - m_{1}\check{p}^{2} - n_{1}\check{p}t - n_{4}\check{p}\check{\mu} \nonumber \\
   &  & +\: O_{3}(t,\check{\mu},\check{p}),  \label{tilde-t} \\
  \tilde{h} & = & \check{\mu} - k_{1}t - j_{2}\check{p} \nonumber \\
   &  & -\: r_{2}t^{2} - q_{2}\check{\mu}^{2} - v_{2}t\check{\mu} - m_{2}\check{p}^{2} - n_{2}\check{p}t - n_{5}\check{p}\check{\mu} \nonumber \\
   &  & +\: O_{3}(t,\check{\mu},\check{\mu}),  \label{tilde-h}
 \end{eqnarray}
where we have introduced the dimensionless critical deviations \cite{fis:ork}
 \begin{equation}
  t \equiv \frac{T-\Tc}{\Tc}, \hspace{0.2in} \check{\mu} = \frac{\mu - \mu_{\mbox{\scriptsize c}}}{k_{\mbox{\scriptsize B}}\Tc}, \hspace{0.2in} \check{p} = \frac{p-p_{\mbox{\scriptsize c}}}{\rhoc k_{\mbox{\scriptsize B}}\Tc},  \label{critical_dev}
 \end{equation}
in which $\rhoc$ is the critical (number) density and $k_{\mbox{\scriptsize B}}$ is Boltzmann's constant, while $O_{m}(x,y,z)$ denotes a formal expansion in powers $\,x^{j}y^{k}z^{l}\,$ with $j+k+l\geq m$. In accepting these expansions we are neglecting any scaling-exponent ``resonances'' that might complicate the expressions by introducing logarithmic factors, etc.\ \cite{wegner,fisher1,fisher2,binney}. For ($d$=3)-dimensional systems, particularly real asymmetric fluids --- which are of especial interest --- this should be quite satisfactory. The irrelevant scaling fields, $u_{4}$, $u_{5}$, $\cdots$, will, in general, have similar expansions in powers of the variables, $t$, $\check{\mu}$ and $\check{p}$. 

As usual, $\alpha$ in (\ref{scaling-descript}) is the universal critical exponent of the specific heat and $\Delta$ is the gap exponent that is related to $\alpha$ and the exponents $\beta$ (for the spontaneous magnetization), $\gamma$ (for the susceptibility) and $\delta$ (for the critical isotherm) by the scaling relations
 \begin{equation}
  \Delta = 2-\alpha-\beta = \beta + \gamma = \beta\delta,  \label{scaling-relation}
 \end{equation}
while $\theta_{4}\equiv\theta$ and $\theta_{5}$ are the {\em positive} leading correction-to-scaling exponents. In the case of the ($d$=3)-dimensional Ising universality class, which is believed to characterize fluid criticality rather generally, we may accept $\alpha \simeq 0.109$, $\beta \simeq0.326$, $\gamma \simeq 1.239$, $\Delta \simeq 1.565$, $\theta \simeq 0.52$ and $\theta_{5} \simeq 1.32$ \cite{kim:fis:bar}. Finally, substituting $\lambda = 1/|\tilde{t}|$ into (\ref{scaling-descript}) yields the ``thermal scaling'' form
 \begin{equation}
  \Phi^{\pm}(\tilde{p}/|\tilde{t}|^{2-\alpha},\:\tilde{h}/|\tilde{t}|^{\Delta};\:u_{4}|\tilde{t}|^{\theta},\:u_{5}|\tilde{t}|^{\theta_{5}},\cdots) = 0, \label{scale}
 \end{equation}
where $\Phi^{\pm}(x,y;\cdots) = \Psi(x,\pm 1,y;\cdots)$ in which $\pm$ corresponds to $\tilde{t}\gtrless 0$.

A further observation is worth mentioning at this point. Specifically, if the critical point is drawn out into a lambda line without changing the universality class, by, e.g., the imposition of a magnetic field, $H$, or, the addition of a second molecular component with chemical potential, say $\mu^{\mbox{\scriptsize II}}$, etc., the only change needed in the formulation is the inclusion of a further, {\em nonordering field} $g~ [~ \propto H~ \mbox{or}~ z_{\mbox{\scriptsize II}}=\exp(-\mu^{\mbox{\scriptsize II}}/k_{\mbox{\scriptsize B}}T)$, etc.] in the expansions of the three nonlinear scaling fields; likewise for any further fields that leave the universality class unaltered: see, e.g., [17,18].

Now the particle number $N$, volume $V$, and the entropy $S$, are related by the Gibbs-Duhem relation
 \begin{equation}
  Vdp - SdT - Nd\mu = 0. \label{gibbs-duhem}
 \end{equation}
Hence the number density and the entropy density are given by
 \begin{equation}
   \rho \equiv \frac{N}{V} = \left( \frac{\partial p}{\partial\mu} \right)_{T}, \hspace{.3in} {\cal S} \equiv \frac{S}{V} = \left( \frac{\partial p}{\partial T} \right)_{\mu}.  \label{rho-s}
 \end{equation}
Other thermodynamic quantities, such as the specific heat, compressibility, etc., follow in the standard way.

The crucial point about the scaling formulation presented here is that the three standard thermodynamic fields, $p$, $\mu$, and $T$ enter in a fully symmetrical way with no {\em a priori} assumptions as to which pair combination or thermodynamic potential, say $p(\mu,T)$ or $\mu(p,T)$, is ``more basic.'' By contrast, previous formulations have typically concluded that it was most appropriate to regard $p(\mu,T)$ as the ``basic'' quantity \cite{mer,kim:fis:bar}. This is what one is led to by studying the standard lattice gases and what arises most naturally from field-theoretic and renormalization group approaches \cite{fisher1,fisher2,binney}. However, one might note that the original Widom formulation \cite{widom} was based on integrating the equation of state and led, fairly naturally, to a scaling hypothesis for the Helmholtz free energy density $f(\rho,T)$; on the other hand, the exactly soluble cluster-interaction fluids of Fisher and Felderhof \cite{fis:fel} showed that $\mu(p,T)$ was the appropriate thermodynamic potential for describing critical-point scaling in these models.

It is easily seen, however, that the previous $p(\mu,T)$ formulation \cite{mer,kim:fis:bar} is recaptured here simply by suppressing the $\check{p}$ dependence in the nonlinear scaling fields $\tilde{h}$ and $\tilde{t}$: i.e., in linear order, by setting $j_{1}=j_{2}=0$ in (\ref{tilde-t}) and (\ref{tilde-h}), and in quadratic order, by setting $m_{0}=m_{1} = \cdots = n_{5}=0$. In that case the expansion of $\tilde{p}$ in (\ref{tilde-g}) merely serves to represent the usual ``smooth background'' $p_{0}(\mu,T)$ that is always required.

The original discussion of scaling in fluids \cite{widom,fisher3} had to incorporate in the definition of the {\em ordering field} $\,\tilde{h}\,$ the {\em mixing coefficient} $\,k_{1}\,$ which, indeed, is then directly proportional to $\,(d\mu_{\sigma}/dT)_{\mbox{\scriptsize c}}\,$ where $\mu_{\sigma}(T)$ represents the phase boundary below $\Tc$; but the coefficients $j_{1}$, $j_{2}$ and $l_{1}$ did not appear. Studies of various models later revealed that the coefficient $l_{1}$, which mixes the chemical potential into the {\em thermal scaling field}, $\tilde{t}$, should also be included: doing so yielded an unexpected singularity in the coexistence curve diameter, $\bar{\rho}(T)$, proportional to $|t|^{1-\alpha}$. (See, e.g.\ Ref.\ [15].) Nevertheless, it was still argued (in Ref.\ [15b]) that one could suppress the linear pressure mixing coefficients $j_{1}$ and $j_{2}$ despite the existence of soluble models in which scaling, in fact, required them \cite{fis:fel}.

It is only the recent reconsideration of the possibility of a Yang-Yang anomaly in real fluids and nonsymmetric models \cite{fis:ork,kos:ani} that has led to the realization that {\em pressure mixing}, i.e., nonvanishing coefficients $j_{1}$ and $j_{2}$ in (\ref{tilde-t}) and (\ref{tilde-h}), should be reconsidered. Indeed, as shown in Ref.\ [2], if a nonvanishing Yang-Yang anomaly arises, so that $(d^{2}\mu_{\sigma}/dT^{2})$ {\em diverges} when $T\rightarrow \Tc-$, then, within a scaling formulation, one must have $j_{2}\neq 0$. In fact, that also leads, as we demonstrate below, to a singular term varying as $|t|^{2\beta}$ in the coexistence curve diameter, $\bar{\rho}(T)$, which, in realistic situations, will actually dominate the previously discovered $|t|^{1-\alpha}$ term \cite{fis:ork,mer}. Conversely, the absence of pressure mixing in the linear ordering field, i.e. $j_{2}=0$, implies the {\em absence} of a Yang-Yang anomaly as is the case for the standard lattice gases. Granted that $j_{2}$ may not vanish, it is clearly appropriate to allow $j_{1}\neq 0$, i.e., to consider pressure mixing also in the thermal field, $\tilde{t}$.

In light of this background the aim of the analysis presented in this paper is to thoroughly explore the implications of the ``complete scaling hypothesis'' embodied in the relations (\ref{scaling-descript})-(\ref{scale}). Hence in Sec.\ \ref{sec2} the ``complete scaling'' theory is formulated with the incorporation of pressure mixing as well as corrections to scaling. For this purpose, we mostly follow the treatment given in Ref.\ [17]. The properties of the scaling functions in the two limiting cases, $\,\tilde{h}/|\tilde{t}|^{\Delta} \rightarrow 0\,$ or $\,\infty,\,$ are set out as well as the consequences of the thermodynamic convexity or the 2nd Law of Thermodynamics \cite{fis:kim}. For use in the subsequent calculations, generalized scaling densities and susceptibilities are defined and the relations between them and the actual physical quantities are determined.

Once the formulation of the scaling theory is completed, we examine, in Sec.\ \ref{sec3}, the standard singular thermodynamic properties including the coexistence curve and the specific heat. We explicitly determine the nature of the phase boundaries, $p_{\sigma}(T)$ and $\mu_{\sigma}(T)$, of the coexistence curve diameter, $\bar{\rho}=\frac{1}{2}(\rho_{\mbox{\scriptsize liq}}+\rho_{\mbox{\scriptsize vap}})$ and of the half-jump, $\Delta\rho = \frac{1}{2}(\rho_{\mbox{\scriptsize liq}}-\rho_{\mbox{\scriptsize vap}})$, as well as of entropies at the coexistence curve. As mentioned above, the pressure-mixing coefficient $j_{2}$ generates a singular $|t|^{2-\alpha}$ term in the phase boundary $\mu_{\sigma}(T)$ and thereby, the divergence of the $(d^{2}\mu_{\sigma}/dT^{2})$ at the critical point: see Sec.\ III.A; in addition, a singular $|t|^{2\beta}$ term in the coexistence curve diameter appears and dominates the $|t|^{1-\alpha}$ term known previously \cite{mer}: see Sec.\ III.B. In Sec.\ III.C the linear mixing coefficients $j_{1}$, $j_{2}$, $\cdots$, $l_{1}$ are related to various physical amplitudes, etc., which can, in principle, be measured via experiments or simulations. The Yang-Yang anomaly, that was the main motivation for this study, is discussed in Sec.\ III.D; it is shown that the strength of the anomaly, namely ${\cal R}_{\mu}$, which measures the contribution of the chemical potential to the heat capacity relative to that of the pressure in an asymptotic limit --- see (\ref{yang-yang}) --- is related solely to the pressure-mixing coefficient $j_{2}$.

In Sec.\ \ref{sec4} we study a number of special loci, particularly in the density-temperature $(\rho,T)$ plane, that intersect the critical point. These loci are of interest in their own right but they prove to be especially useful in attempting to estimate precise values of the critical density, $\rhoc$, both in the case of real asymmetric fluids and in the simulation of various model systems such as the hard-core square-well fluid \cite{ork:fis:pan} and, more challengingly, primitive model electrolyte systems \cite{luijten2}. More concretely, we study the locus $\mu = \mu_{\mbox{\scriptsize c}}$, dubbed the {\em critical isokyme}, the critical isobar $(p=p_{\mbox{\scriptsize c}})$, the critical isotherm $(T=\Tc)$ and the critical isochore $(\rho=\rhoc)$ in the $(p,T)$, $(\mu,T)$, $(\rho,T)$, etc., planes. The singular exponents characterizing these loci are obtained and the corresponding amplitudes are expressed in terms of the mixing coefficients and the expansion coefficients of the scaling functions and, thereby, related to one another. One discovers, however, that these characterizations of the various loci may not be helpful in estimating the critical point in practical applications to experiments or simulations, since various critical parameter values must be known {\em a priori}; but, except for exactly soluble models, the requisite values are not normally available.

Hence, for direct applications to the analysis of simulation data, we also study in Sec.\ IV.E, the ``$k$-inflection loci'' introduced by Orkoulas {\em et al.} \cite{ork:fis:pan}. The ``$k$-susceptibility loci'' are defined by the points of isothermal maxima of the modified susceptibility $\chi^{(k)}=\chi /\rho^{k}$ above $T_{\mbox{\scriptsize c}}$ in the $(\rho,T)$ plane where $\chi = \rho^{2}k_{\mbox{\scriptsize B}}T K_{T}$ is the standard isothermal susceptibility while $K_{T}$ is the compressibility. We find that these loci have leading $|t|^{2\beta}$ terms followed by more slowly varying $|t|^{1-\alpha}$ and $t$ contributions: however, the leading amplitude {\em vanishes} when $k$ takes some `optimal value' $k_{\mbox{\scriptsize opt}}$. Thus, when $k=k_{\mbox{\scriptsize opt}}$, the corresponding $k$-locus ``points'' most directly to the critical point. Furthermore, we find that $k_{\mbox{\scriptsize opt}}$ is directly related to the strength of the Yang-Yang anomaly via $k_{\mbox{\scriptsize opt}}=3{\cal R}_{\mu}$. With the aid of simulations we illustrate these loci quantitatively for the hard-core square-well fluid and the restricted primitive model electrolyte, and compare the results with those for a van der Waals fluid: see Figs.\ \ref{fig1}-\ref{fig3}, below.

One may, similarly \cite{yckim}, define ``$k$-heat-capacity loci'' via the points of isothermal maxima in the $(\rho,T)$ plane of the modified specific heat $C_{V}^{(k)}\equiv C_{V}/\rho^{k}$. These loci have not been used in past simulations but may be useful in the future. We find, as for the susceptibility loci, that a leading $|t|^{2\beta}$ term appears in the near-critical expansions, followed by $|t|^{1-\alpha}$ and $t$ terms. Again, the leading amplitude vanishes when $k$ is equal to a special value that is once more related directly to the Yang-Yang anomaly: see Sec.\ IV.F. 

In order to provide semiquantitative guides to the behavior of these loci in real systems, we study them in Sec.\ \ref{sec5}, using classical theory for the gas-liquid phase transition on the basis of a Landau order-parameter expansion of the free energy. We obtain the vapor pressure, $p_{\sigma}(T)$, and the saturation chemical potential, $\mu_{\sigma}(T)$, up to second order in $t$, while in the current literature they appear only up to first order \cite{sengers}. The coexistence curve diameter, $\bar{\rho}(T)$, is also found to order $t$. These curves can be analytically continued from $T \leq T_{\mbox{\scriptsize c}}$ to the one-phase region above $T_{\mbox{\scriptsize c}}$ so providing special loci that cannot, in general, be defined in nonclassical cases. Using the van der Waals equation, these analytic extensions are illustrated quantitatively: see Figs.\ \ref{fig5}-\ref{fig7}. In addition to these critical loci, the $k$-susceptibility-loci are obtained and compared with those of the hard-core square-well  \cite{ork:fis:pan} and the restricted primitive model \cite{luijten2}. In Sec.\ V.D the Yang-Yang relation (\ref{yang-yang}) is discussed in more detail within classical theory and extended to general loci in the $(\rho,T)$ plane. Finally, a Yang-Yang-type relation along the critical isotherm is discussed briefly.

Sec.\ \ref{sec6} summarizes the paper and provides a key to the main results. Some further details are provided in the Appendix and others are available in the first author's thesis \cite{yckim}.

\section{Scaling Formulation and Thermodynamic Functions}
\label{sec2}
To explore the scaling description (\ref{scale}), it is appropriate to focus on the relevant scaling variable (or combination)
 \begin{equation}
  y(\check{p},\check{\mu},t) = U\tilde{h}/|\tilde{t}|^{\Delta},  \label{y}
 \end{equation}
where, without loss of generality, $U$ may be taken as a positive constant. The basic reason for this choice, as against $\tilde{p}/|\tilde{t}|^{2-\alpha}$, is that $\Delta$ is, in general, less than $(2-\alpha)$ [since, see (\ref{scaling-relation}), $\beta > 0$] so that $y$ diverges less rapidly when $t\rightarrow 0$. Beyond $y$ we need to account for the full set of irrelevant scaled variables, namely,
 \begin{eqnarray}
   y_{k}(\check{p},\check{\mu},t) & = & U_{k}(\check{p},\check{\mu},t)|\tilde{t}|^{\theta_{k}}, \nonumber \\
   &  & \theta_{k+1} \geq \theta_{k} >0, \hspace{.2in} k = 4,5,\cdots,  \label{yk}
 \end{eqnarray}
where $U_{k} \propto u_{k}$ and we will assume, when needed, that the associated irrelevant amplitudes $U_{k}$ are noncritical, meaning that $U_{k}(\check{p},\check{\mu},t)$ can be expanded in a formal power series of $\check{p}$, $\check{\mu}$ and $t$ (which may not necessarily converge). Since a fluid has, in general, no obvious symmetry, we do {\em not} impose any restriction on the $U_{k}$: but see also the discussion in Ref.\ [17] Sec.\ II.

Now the thermodynamic potential for the fluid (in this case the pressure --- as a consequence of selecting $y$ as the primary scaled variable) can be written by formally solving (\ref{scale}) as
 \begin{equation}
  \tilde{p} = Q|\tilde{t}|^{2-\alpha} W_{\pm}(y,y_{4},y_{5},\cdots), \label{scaling-eq}
 \end{equation}
where $Q$ is a positive amplitude while $W_{\pm}$ is a scaling function that, in the case of bulk fluids, embodies the properties of the ($d=3$)-dimensional Ising universality class. In this expression the subscript $\pm$ refers to $\tilde{t} \gtrless 0$. On choosing appropriate values for $Q$ and $U$, the scaling function, $W_{\pm}(y,\cdots)$ becomes universal. Note that the usual analytic background part of the potential, say $p_{0}(t,\check{\mu})$, is included in the nonlinear scaling field $\tilde{p}$: see (\ref{tilde-g}) which may, clearly, be solved iteratively for $\check{p}$ to yield an expansion for $p_{0}(t,\check{\mu})$.

\vspace{0.15in}
\subsection{Scaling functions}
\label{sec2.1}
The scaling function $W_{\pm}(y,y_{4},y_{5},\cdots)$ should be both universal and invariant under change of sign of the odd arguments $y$, $y_{5}$, $y_{7}$, $\cdots$. Since the irrelevant scaling variables, $y_{4}$, $y_{5}$, $\cdots$, vanish as the critical point is approached, we can also expand the scaling function $W_{\pm}$ as \cite{kim:fis:bar}
 \begin{eqnarray}
  & &\hspace{-0.4in} W_{\pm}(y,y_{4},y_{5},\cdots) \nonumber \\
  & = & W_{\pm}^{0}(y) + y_{4}W_{\pm}^{(4)}(y) + y_{5}W_{\pm}^{(5)}(y) + \cdots \nonumber \\
  &  &  +\: y_{4}^{2}W_{\pm}^{(4,4)}(y) + y_{4}y_{5}W_{\pm}^{(4,5)}(y) + \cdots \nonumber \\
  & = &  \sum_{\mbox{\scriptsize \boldmath $\kappa$}}W_{\pm}^{\mbox{\scriptsize \boldmath $\kappa$}}(y) y^{[\mbox{\scriptsize \boldmath $\kappa$}]},   \label{Wpm}
 \end{eqnarray}
where for brevity we have used the multi-index, {\boldmath $\kappa$}, defined via
 \begin{equation}
  \mbox{\boldmath $\kappa$} = 0, (4), (5), \cdots, (4,4), (4,5), \cdots, (4,4,4), \cdots,  \label{kappa}
 \end{equation}
as a label and as an exponent via
 \begin{equation}
  y^{0} \equiv 1, \hspace{0.2in} y^{[i,j,\cdots,n]} \equiv y_{i}y_{j}\cdots y_{n}.  \label{yijn}
 \end{equation}
We consider $\mbox{\boldmath $\kappa$}=(i,j,\cdots,n)$ to be odd or even according to whether the sum $\,i+j+\cdots +n\,$ is odd or even. The symmetry of $\,W_{\pm}(y,y_{4},y_{5},\cdots)$ then requires \cite{kim:fis:bar}
 \begin{equation}
  W_{\pm}^{\mbox{\scriptsize \boldmath $\kappa$}}(-y) = (-)^{\mbox{\scriptsize \boldmath $\kappa$}} W_{\pm}^{\mbox{\scriptsize \boldmath $\kappa$}}(y).  \label{Wpm-symmetry}
 \end{equation}

Recognizing this symmetry, one may write expansions for $W_{\pm}^{\mbox{\scriptsize \boldmath $\kappa$}}(y)$ for small $y$. For $\tilde{t}>0$ we have
 \begin{eqnarray}
 & &\hspace{-0.2in}  W_{+}^{\mbox{\scriptsize \boldmath $\kappa$}}(y) \nonumber \\
 & = & W_{+0}^{\mbox{\scriptsize \boldmath $\kappa$}} + y^{2} W_{+2}^{\mbox{\scriptsize \boldmath $\kappa$}} + y^{4} W_{+4}^{\mbox{\scriptsize \boldmath $\kappa$}} + \cdots, \hspace{0.1in} \mbox{for  \boldmath $\kappa$ even},  \nonumber \\
  & = & y W_{+1}^{\mbox{\scriptsize \boldmath $\kappa$}} + y^{3} W_{+3}^{\mbox{\scriptsize \boldmath $\kappa$}} + y^{5} W_{+5}^{\mbox{\scriptsize \boldmath $\kappa$}} + \cdots, \hspace{0.1in} \mbox{for  \boldmath $\kappa$ odd}.  \label{Wp-expan}
 \end{eqnarray}
By choosing appropriate values for the nonuniversal metric amplitudes $Q$, $U$, etc., the expansion coefficients can be normalized, in general, such that \cite{kim:fis:bar,fis:kim}
 \begin{equation}
  W_{+2}^{0} = W_{+0}^{\mbox{\scriptsize \boldmath $\kappa$}} = 1 \hspace{0.1in} \mbox{(\boldmath $\kappa$  even)} \hspace{0.1in} \mbox{or} \hspace{0.1in} W_{+1}^{\mbox{\scriptsize \boldmath $\kappa$}} = 1 \hspace{0.1in} \mbox{(\boldmath $\kappa$ odd)}.  \label{coeff}
 \end{equation}

For $\tilde{t} < 0$, the existence of the first-order transition leads to $|y|$ factors in the expansions so that one has \cite{kim:fis:bar}
 \begin{equation}
  W_{-}^{\mbox{\scriptsize \boldmath $\kappa$}}(y) = [W_{-0}^{\mbox{\scriptsize \boldmath $\kappa$}} + |y| W_{-1}^{\mbox{\scriptsize \boldmath $\kappa$}} + y^{2} W_{-2}^{\mbox{\scriptsize \boldmath $\kappa$}} + |y|^{3} W_{-3}^{\mbox{\scriptsize \boldmath $\kappa$}} + \cdots ] \sigma_{\mbox{\scriptsize \boldmath $\kappa$}}(y),  \label{Wm-expan}
 \end{equation}
where the special signum function is defined by
 \begin{eqnarray}
  \sigma_{\mbox{\scriptsize \boldmath $\kappa$}}(y) & = & 1 \hspace{0.55in} \mbox{for  \boldmath $\kappa$ even,}  \nonumber \\
  & = & \mbox{sgn}(y) \hspace{0.2in} \mbox{for \boldmath $\kappa$ odd.} \label{signum}
 \end{eqnarray}
Thermodynamic convexity (which embodies the Second Law) then requires that $W_{-1}^{0}$ and $W_{-2}^{0}$ must both be positive: see Ref.\ \cite{fis:kim}.

For large arguments, $|y|\rightarrow \infty$, the scaling functions $W_{+}^{\mbox{\scriptsize \boldmath $\kappa$}}(y)$ and $W_{-}^{\mbox{\scriptsize \boldmath $\kappa$}}(y)$ must satisfy stringent matching conditions to ensure the analyticity of the potential through the plane $\tilde{t} = 0$ for all $\tilde{h}\neq 0$. By these conditions, the functions $W_{\pm}^{\mbox{\scriptsize \boldmath $\kappa$}}(y)$ may be written as
 \begin{eqnarray}
  W_{\pm}^{\mbox{\scriptsize\boldmath $\kappa$}}(y) & \approx & W_{\infty}^{\mbox{\scriptsize \boldmath $\kappa$}}|y|^{(2-\alpha + \theta[\mbox{\scriptsize \boldmath $\kappa$}])/\Delta} \nonumber \\
  &  & \times \left[ 1 + \sum_{l=1}^{\infty} (\pm)^{l} w_{l}^{\mbox{\scriptsize \boldmath $\kappa$}} |y|^{-l/\Delta} \right] \sigma_{\mbox{\scriptsize \boldmath $\kappa$}}(y),  \label{Wpm-highexpan}
 \end{eqnarray}
where the multiexponent $\theta[\mbox{\boldmath $\kappa$}]$ is defined by
 \begin{equation}
  \theta[0] \equiv 0, \hspace{0.1in} \theta[(i,j,\cdots,n)] = \theta_{i} + \theta_{j} + \cdots + \theta_{n},  \label{theta-kappa}
 \end{equation}
with $i,\, j,\, \cdots,\, n \geq 4$. By virtue of the normalizations (\ref{coeff}) the numerical amplitudes $W_{-j}^{\mbox{\scriptsize \boldmath $\kappa$}}$, $W_{+j}^{\mbox{\scriptsize \boldmath $\kappa$}}$, $W_{\infty}^{\mbox{\scriptsize \boldmath $\kappa$}}$, and $w_{l}^{\mbox{\scriptsize \boldmath $\kappa$}}$ should all be universal. Beyond that, thermodynamic convexity dictates that $W_{\infty}^{0}$ and $w_{2}^{0}$ must be positive while $(w_{1}^{0})^{2}/w_{2}^{0}$ must be bounded above \cite{fis:kim}. The sign of $w_{1}^{0}$ is not determined by convexity alone but must, in general, be negative: see Ref.\ \cite{fis:kim}. This plays an important role in determining allowable phase diagrams in a density (or composition) space at critical endpoints \cite{kim:fis:bar,fis:kim}.

\subsection{Generalized density and entropy}
\label{sec2.2}
To extract explicit results from (\ref{scaling-eq}) revealing the singularities of the various thermodynamic derivatives, critical loci, etc., we first recall (\ref{rho-s}) and define a dimensionless reduced density, $\check{\rho}$, and entropy, $\check{s}$, by
 \begin{equation}
  \check{\rho} \equiv \frac{\rho}{\rhoc} = \left( \frac{\partial \check{p}}{\partial \check{\mu}} \right)_{t}, \hspace{0.1in} \check{s} \equiv \frac{\cal S}{\rhoc k_{\mbox{\scriptsize B}}} = \left( \frac{\partial \check{p}}{\partial t} \right)_{\check{\mu}}. \label{red-den}
 \end{equation}
Similarly, the generalized number density, $\tilde{\rho}$, and entropy density, $\tilde{s}$, will be defined by
 \begin{equation}
  \tilde{\rho} \equiv \left( \frac{\partial \tilde{p}}{\partial \tilde{h}}\right)_{\tilde{t}}, \hspace{0.2in}  \tilde{s} \equiv \left( \frac{\partial\tilde{p}}{\partial\tilde{t}} \right)_{\tilde{h}}.  \label{gen-den}
 \end{equation}
These quantities then obey the usual simple scaling rules, namely, $\tilde{\rho} \sim |\tilde{t}|^{\beta}$ and $\tilde{s} \sim |\tilde{t}|^{1-\alpha}$ when $\tilde{t}\rightarrow 0$.

To find concrete expressions for $\check{\rho}$ and $\check{s}$, one may consider the differential relation,
 \begin{equation}
  d\tilde{p} = \tilde{\rho} d\tilde{h} + \tilde{s} d\tilde{t}, \label{diff-rel}
 \end{equation}
where it is appropriate to recall that all the scaling variables $y_{k}$ in (\ref{scaling-eq}) are functions (only) of $\check{p}$, $\check{\mu}$ and $t$ and, hence, of $\tilde{p}$, $\tilde{\mu}$ and $\tilde{t}$. Using (\ref{tilde-g})-(\ref{tilde-h}), this differential relation may be written as
 \begin{eqnarray}
  (1 & - & 2m_{0}\check{p} - n_{0}\check{\mu} -n_{3}t + j_{2}\tilde{\rho} + j_{1}\tilde{s}+ \cdots)d\check{p} \nonumber \\
  & =  & (l_{0} + n_{0}\check{p} + 2q_{0}\check{\mu} + v_{0}t + \tilde{\rho} - l_{1}\tilde{s} + \cdots)d\check{\mu} \nonumber \\
  &   & \hspace{0.1in} +\: (k_{0} n_{3}\check{p} + v_{0}\check{\mu} + 2r_{0}t + \tilde{s} - k_{1}\tilde{\rho} + \cdots)dt, \label{diff-rel2}
 \end{eqnarray}
where in the brackets we have retained only terms up to linear order. Using (\ref{red-den}), we then obtain the relations
 \begin{eqnarray}
  \check{\rho} & = & \frac{l_{0} + n_{0}\check{p} + 2q_{0}\check{\mu} + v_{0}t + \tilde{\rho} - l_{1}\tilde{s} + \cdots}{1 - 2m_{0}\check{p} - n_{0}\check{\mu} - n_{3}t + j_{2}\tilde{\rho}+j_{1}\tilde{s} + \cdots},  \label{checkrho-eq1} \\
  \check{s} & = & \frac{k_{0} + n_{3}\check{p} + v_{0}\check{\mu} + 2r_{0}t + \tilde{s} - k_{1}\tilde{\rho}}{1 - 2m_{0}\check{p} - n_{0}\check{\mu} - n_{3}t + j_{2}\tilde{\rho}+j_{1}\tilde{s} + \cdots}.  \label{checks-eq1}
 \end{eqnarray}
If the {\em nonlinear mixing coefficients} $m_{j}$, $n_{j}$, $\cdots$, $v_{j}$ in (\ref{tilde-g})-(\ref{tilde-h}) are ignored, these expressions may be approximated by
 \begin{equation}
  \check{\rho} \approx \frac{l_{0} + \tilde{\rho} - l_{1}\tilde{s}} {1 + j_{2}\tilde{\rho} + j_{1}\tilde{s}}, \hspace{0.3in} \check{s} \approx \frac{k_{0} + \tilde{s} - k_{1}\tilde{\rho}}{1 + j_{2}\tilde{\rho} + j_{1}\tilde{s}}. \label{rel1}
 \end{equation}
Note that $\check{\rho}$ and $\check{s}$ are {\em nonlinear} functions of $\tilde{\rho}$ and $\tilde{s}$ owing to the pressure mixing coefficients, $j_{1}$ and $j_{2}$. In the absence of pressure mixing ({\em i.e.}, $j_{1}=j_{2}=0$), we recover the Bruce-Wilding {\em linear} relations \cite{bruce:wilding}. We will see later that the approximation (\ref{rel1}) is adequate to derive the most important singularities in leading order.

\subsection{Generalized susceptibilities}
\label{sec2.3}
The second derivatives of the potential determine the usual response functions, e.g., susceptibilities, heat capacities, etc. We define the basic reduced susceptibilites
 \begin{eqnarray}
  \check{\chi}_{NN} & \equiv & (\partial^{2}\check{p}/\partial\check{\mu}^{2})_{t}, \hspace{0.2in} \check{\chi}_{UU} \equiv (\partial^{2}\check{p}/\partial t^{2})_{\check{\mu}}, \nonumber \\
  \check{\chi}_{NU} & \equiv & (\partial^{2}\check{p}/\partial\check{\mu}\partial t) = (\partial^{2}\check{p}/\partial t \partial\check{\mu}). \label{suscept}
 \end{eqnarray}
These are related to the number and energy fluctuations most directly accessible in grand canonical simulations \cite{ork:fis:pan}. The isothermal susceptibility $\chi$ is defined by
 \begin{equation}
  \chi \equiv (\partial\rho/\partial\mu)_{T} = (\partial^{2} p/\partial\mu^{2})_{T},  \label{chi}
 \end{equation}
and is related to the isothermal compressibility, $K_{T} = \rho^{-1}(\partial\rho/\partial p)_{T}$, by $\chi = \rho^{2} K_{T}$ and to the reduced susceptibility $\check{\chi}_{NN}$ by
 \begin{equation}
  \chi = (\rhoc/k_{\mbox{\scriptsize B}}\Tc)\check{\chi}_{NN}.  \label{chi_chiNN}
 \end{equation}

Experimentally, the most important heat capacity for fluids is the constant-volume heat capacity with a density defined by
 \begin{equation}
   C_{V} = \frac{T}{\rho}\left( \frac{\partial \cal S}{\partial T} \right)_{\rho}, \label{Cv}
 \end{equation}
where ${\cal S}$ is the entropy density defined in (\ref{rho-s}). It is convenient to introduce the dimensionless reduced specific heat, namely, 
 \begin{equation}
  \check{C}_{V} \equiv (\partial \check{s}/\partial t)_{\rho},  \label{red-sp}
 \end{equation}
for which one has
 \begin{equation}
  \rho C_{V}/T = (k_{\mbox{\scriptsize B}}\rhoc/\Tc)\check{C}_{V}. \label{Cv-check_Cv}
 \end{equation}
 The reduced specific heat, $\check{C}_{V}$, is then related to the reduced susceptibilities in (\ref{suscept}) via \cite{ork:fis:pan}
 \begin{equation}
  \check{C}_{V} = \check{\chi}_{UU} - \check{\chi}_{NU}^{2}/\check{\chi}_{NN}.  \label{red-sp2}
 \end{equation}

Finally, we define generalized (scaling) susceptibilities via
 \begin{equation}
  \tilde{\chi}_{hh} \equiv \left( \frac{\partial^{2}\tilde{p}}{\partial\tilde{h}^{2}}\right)_{\tilde{t}}, \hspace{0.1in} \tilde{\chi}_{tt} \equiv \left(\frac{\partial^{2}\tilde{p}}{\partial\tilde{t}^{2}}\right)_{\tilde{h}}, \hspace{0.1in} \tilde{\chi}_{ht} \equiv \left(\frac{\partial^{2}\tilde{p}}{\partial\tilde{h}\partial{\tilde{t}}}\right).   \label{gen-suscept}
 \end{equation}
From (\ref{scaling-eq}) one finds $\,\tilde{\chi}_{hh}\sim |\tilde{t}|^{-\gamma}$, $\,\tilde{\chi}_{ht} \sim |\tilde{t}|^{\beta -1},\,$ and $\,\tilde{\chi}_{tt} \sim |\tilde{t}|^{-\alpha}\,$ when $\tilde{t}\rightarrow 0$. When the nonlinear mixing terms in the scaling fields are ignored, as above, $\check{\chi}_{NN}$ can be expressed in terms of the generalized densities and susceptibilities as
 \begin{eqnarray}
  \check{\chi}_{NN} & \approx & \left[ (e_{1} + e_{2}\tilde{s})^{2}\tilde{\chi}_{hh} + (e_{3} + e_{2}\tilde{\rho})^{2}\tilde{\chi}_{tt} -2(e_{1} + e_{2}\tilde{s}) \right. \nonumber \\
  &  & \left. \times (e_{3} + e_{2}\tilde{\rho})\tilde{\chi}_{ht}\right]/(1+j_{2}\tilde{\rho}+j_{1}\tilde{s})^{3},  \label{chi-gen}
 \end{eqnarray}
where the derived mixing coefficients are
 \begin{equation}
  e_{1} = 1 - j_{2}l_{0}, \hspace{0.1in} e_{2} = j_{1} + j_{2}l_{1}, \hspace{0.1in} e_{3} = l_{1} + j_{1}l_{0}.  \label{chi-gen-coeff}
 \end{equation}
The detailed derivation of (\ref{chi-gen}) is presented in Appendix F of Ref.\ [26] henceforth to be denoted {\bf K}; there it is also shown how (\ref{chi-gen}) [{\bf K}(3.41)] captures the leading singular correction terms.

\section{Thermodynamic Properties}
\label{sec3}
\subsection{Phase boundaries}
\label{sec3.1}
The phase boundaries, say $p_{\sigma}(T)$ and $\mu_{\sigma}(T)$ [or $\check{p}_{\sigma}(t)$ and $\check{\mu}_{\sigma}(t)$], on which two phases may coexist, can be determined by equating the pressure and chemical potential on the vapor and liquid sides below $\Tc$ ({\em i.e.,} for $\tilde{t}<0$). On using the small $y$ expansion of $\tilde{p}$ for $\tilde{t}<0$ with the aid of (\ref{Wm-expan}), we obtain (including higher order terms)
 \begin{eqnarray}
  \tilde{p}_{\pm} & = & Q|\tilde{t}|^{2-\alpha} \mbox{\LARGE $[$} W_{-0}^{0} \pm W_{-1}^{0}y + W_{-2}^{0}y^{2} + \cdots \nonumber \\
   &  & +\: U_{4\mbox{\scriptsize c}}|\tilde{t}|^{\theta}\mbox{\Large $($}W_{-0}^{(4)} \pm W_{-1}^{(4)}y + W_{-2}^{(4)}y^{2} + \cdots\mbox{\Large $)$}  \nonumber \\
  &  &  \hspace{0.in} \pm \: U_{5\mbox{\scriptsize c}}|\tilde{t}|^{\theta_{5}}\mbox{\Large $($}W_{-0}^{(5)} \pm W_{-1}^{(5)}y + W_{-2}^{(5)}y^{2} + \cdots\mbox{\Large $)$} + \cdots\mbox{\LARGE $]$},  \label{boundary-cond}
 \end{eqnarray}
where $\pm$ now refers to $\tilde{h} \gtrless 0$, while $U_{4\mbox{\scriptsize c}}$ and $U_{5\mbox{\scriptsize c}}$ are the critical values of the irrelevant scaling amplitudes, $U_{4}$ and $U_{5}$, respectively. Equating $\tilde{p}_{+}$ and $\tilde{p}_{-}$ then yields
 \begin{eqnarray}
  W_{-1}^{0}y & = & -\: W_{-0}^{(5)}U_{5\mbox{\scriptsize c}}|\tilde{t}|^{\theta_{5}} - W_{-1}^{(4)}U_{4\mbox{\scriptsize c}}|\tilde{t}|^{\theta}y \nonumber \\
  &  & -\: W_{-2}^{(5)}U_{5\mbox{\scriptsize c}}|\tilde{t}|^{\theta_{5}}y^{2} + \cdots,  \label {y-boundary}
 \end{eqnarray}
where one sees that various even and odd terms cancel.
Solving for $y$ iteratively, one obtains
 \begin{equation}
  y = -\: \frac{W_{-0}^{(5)}}{W_{-1}^{0}}U_{5\mbox{\scriptsize c}}|\tilde{t}|^{\theta_{5}} + \frac{W_{-1}^{(4)}W_{-0}^{(5)}}{(W_{-1}^{0})^{2}} U_{4\mbox{\scriptsize c}}U_{5\mbox{\scriptsize c}}|\tilde{t}|^{\theta+\theta_{5}} + \cdots.  \label{y-sol}
 \end{equation}
Note that in contrast to the symmetric case (e.g., the ferromagnetic Ising model) where $U_{5}=0$, the scaling field does {\em not} vanish identically along the phase boundary.
Using the definition (\ref{y}), we find the phase boundary is given by
 \begin{equation}
  \tilde{h}_{\sigma}(\tilde{t}) = E_{1} |\tilde{t}_{\sigma}|^{\Delta + \theta_{5}} + E_{2} |\tilde{t}_{\sigma}|^{\Delta + \theta + \theta_{5}} + \cdots,  \label{tilde-h-sigma}
 \end{equation}
where the coefficients are
 \begin{equation}
  E_{1} = -\: \frac{U_{5\mbox{\scriptsize c}} W_{-0}^{(5)}}{UW_{-1}^{0}}, \hspace{0.1in} E_{2} = \frac{U_{4\mbox{\scriptsize c}}U_{5\mbox{\scriptsize c}}W_{-1}^{(4)}W_{-0}^{(5)}}{U (W_{-1}^{0})^{2}}.  \label{h-coeff}
 \end{equation}

Now substituting $\mu = \mu_{\sigma}(T)$ and $p = p_{\sigma}(T)$ in (\ref{tilde-t}) and (\ref{tilde-h}) and using (\ref{tilde-h-sigma}), we obtain an expansion [see {\bf K}(3.65)] of
 \begin{equation}
  \check{\mu}_{\sigma}(T) = [\mu_{\sigma}(T)-\mu_{\mbox{\scriptsize c}}]/k_{\mbox{\scriptsize B}}T_{\mbox{\scriptsize c}} \label{boundary-a}
 \end{equation}
in powers of $t$, of $\check{\mu}_{\sigma}$ itself, and of
 \begin{equation}
  \check{p}_{\sigma}(T) = [p_{\sigma}(T)-p_{\mbox{\scriptsize c}}]/\rho_{\mbox{\scriptsize c}}k_{\mbox{\scriptsize B}}T_{\mbox{\scriptsize c}}. \label{boundary}
 \end{equation}
One can solve this equation iteratively for $\check{\mu}_{\sigma}$ as a function of $t$ and $\check{p}_{\sigma}$ to obtain
 \begin{eqnarray}
  \check{\mu}_{\sigma}(T) & = & j_{2}\check{p}_{\sigma} + k_{1}t + (m_{2}+j_{2}^{2}q_{2}+j_{2}n_{2})\check{p}_{\sigma}^{2} \nonumber \\
  &  & +\: (r_{2}+k_{1}q_{2}+v_{2}k_{1})t^{2}  \nonumber \\
  &  & +\: (n_{5}+2j_{2}k_{1}q_{2} + k_{1}n_{2} + v_{2}j_{2})\check{p}_{\sigma}t \nonumber \\
  &  & +\: E_{1}\mbox{\Large $|$}(1-k_{1}l_{1})t - (j_{1}+j_{2}l_{1})\check{p}_{\sigma}\mbox{\Large $|$}^{\Delta + \theta_{5}} \nonumber \\
  &  & +\: E_{2}\mbox{\Large $|$}(1-k_{1}l_{1})t - (j_{1}+j_{2}l_{1})\check{p}_{\sigma}\mbox{\Large $|$}^{\Delta + \theta + \theta_{5}} \nonumber \\
  &  & + \cdots.  \label{mu-sigma}
 \end{eqnarray}
Substituting this result into (\ref{tilde-g}) yields
 \begin{eqnarray}
  \tilde{p}_{\sigma} & = & (1-j_{2}l_{0})\check{p}_{\sigma} - (k_{0}+k_{1}l_{0})t \nonumber \\
  &  & -\: \tilde{E}_{1}\check{p}_{\sigma}^{2} - \tilde{E}_{2}t^{2} - \tilde{E}_{3}\check{p}_{\sigma}t + \cdots,  \label{tilde-g-sigma}
 \end{eqnarray}
where the coefficients $\tilde{E}_{j}$ depend on the {\em quadratic} nonlinear-field coefficients $r_{0}$, $m_{0}$, $\cdots$ in (\ref{tilde-g}) to (\ref{tilde-h}): see {\bf K}(3.68)-(3.70).

Similarly, from (\ref{tilde-t}) and (\ref{mu-sigma}), one finds
 \begin{equation}
  \tilde{t}_{\sigma} = (1-k_{1}l_{1})t - (j_{1}+j_{2}l_{1})\check{p}_{\sigma} + \cdots.  \label{tilde-t-sigma}
 \end{equation}

Substituting this into (\ref{boundary-cond}) with the aid of (\ref{y}) and (\ref{tilde-h-sigma}), combining it with (\ref{tilde-g-sigma}) and solving iteratively for $\check{p}_{\sigma}$ finally yields the result
 \begin{eqnarray}
  \check{p}_{\sigma}(T) & = & \check{p}_{\sigma,1}t + \check{p}_{\sigma,2}t^{2} + A_{p}|t|^{2-\alpha}\mbox{\Large $[$} 1 + a_{p}|t|^{\theta} + \cdots \nonumber \\
  &  & +\: b_{p}|t|^{\theta_{5}-\beta} + a_{p}^{\prime}|t|^{\theta_{5}-\beta+\theta} + \cdots + b_{p}^{\prime}|t|^{2\theta_{5}} + \cdots \mbox{\Large $]$} \nonumber \\
  &  & + \cdots, \label{p-boundary}
 \end{eqnarray}
where the leading coefficients are given by
 \begin{eqnarray}
  \check{p}_{\sigma,1} & = &\frac{k_{0} + k_{1}l_{0}}{1 - j_{2}l_{0}}, \hspace{0.1in} \check{p}_{\sigma,2} = \tilde{E}_{1}\check{p}_{\sigma,1}^{2} + \tilde{E}_{2} + \tilde{E}_{3}\check{p}_{\sigma,1}, \label{p-coeff-a} \\
   A_{p} & = & \frac{Q W_{-0}^{0}}{1-j_{2}l_{0}}|\tau|^{2-\alpha}, \hspace{0.1in}  a_{p} = \frac{U_{4\mbox{\scriptsize c}}W_{-0}^{(4)}}{W_{-0}^{0}}|\tau|^{\theta}, \nonumber \\
  b_{p} & = & \frac{l_{0}(1-j_{2}l_{0})E_{1}}{QW_{-0}^{0}}|\tau|^{\theta_{5}-\beta},\label{p-coeff}
 \end{eqnarray}
in which we have introduced mixing factor
 \begin{equation}
  \tau = 1-k_{1}l_{1} - \check{p}_{\sigma,1}(j_{1} + j_{2}l_{1}), \label{tau1}
 \end{equation}
while $E_{1}$ is given in (\ref{h-coeff}) and $a_{p}^{\prime}$ and $b_{p}^{\prime}$ can be derived from {\bf K}(3.74).

Note that the leading even correction-to-scaling term enters the phase boundary $p_{\sigma}(T)$ with an exponent $(2-\alpha)+\theta$, while the odd correction-to-scaling term has an exponent $(2-\alpha)-\beta+\theta_{5}$. In addition, it is not hard to see that the subsequent term $b_{p}|t|^{2\theta_{5}}$ (and analogous terms below) {\em must} be preceded by {\em lower order} terms such as $d_{p}|t|^{2\theta}$, $a_{p}^{\prime\prime}|t|^{\theta +1}$, $e_{p}|t|^{3\theta}$, $a_{p}^{\prime\prime\prime}|t|^{\theta+2}$, etc.

Now let us substitute the result (\ref{p-boundary}) back into (\ref{mu-sigma}). With a little further effort, one obtains the desired result for the chemical potential on the phase boundary, namely,
 \begin{eqnarray}
  \check{\mu}_{\sigma}(T) & = &  \check{\mu}_{\sigma,1}t + \check{\mu}_{\sigma,2}t^{2} + A_{\mu}|t|^{2-\alpha}\mbox{\Large $[$} 1+ a_{\mu}|t|^{\theta} + \cdots \nonumber \\
  &  & +\: b_{\mu}|t|^{\theta_{5}-\beta} + a_{\mu}^{\prime}|t|^{\theta_{5}-\beta+\theta} + \cdots + b_{\mu}^{\prime}|t|^{2\theta_{5}} + \cdots \mbox{\Large $]$} \nonumber \\
  &  & + \cdots,  \label{mu-boundary}
 \end{eqnarray}
where the leading coefficients satisfy
  \begin{eqnarray}
   \check{\mu}_{\sigma,1} & = & k_{1} + j_{2}\check{p}_{\sigma,1}, \hspace{0.1in} A_{\mu} = j_{2}A_{p}, \hspace{0.1in} a_{\mu} = a_{p}, \label{mu-coeff-a} \\
     b_{\mu} & = & E_{1}|\tau|^{\Delta+\theta_{5}}/A_{\mu},\hspace{0.1in}  a_{\mu}^{\prime} = b_{\mu}c_{p}, \hspace{0.1in} b_{\mu}^{\prime} = b_{p}^{\prime},  \label{mu-coeff}
 \end{eqnarray}
while $\check{\mu}_{\sigma,2}$ is given in {\bf K}(3.78).

As is to be expected, the spectrum of singular terms that appears in the expansion for $\mu_{\sigma}(T)$ is the same as that for $p_{\sigma}(T)$. Note, however, that the leading singular amplitude, $A_{\mu}$, for $\mu_{\sigma}(T)$ is proportional to the pressure-mixing coefficient $j_{2}$. Thus, in contrast to the traditional scaling treatment, pressure mixing in the scaling fields implies that the second derivative of $\mu_{\sigma}(T)$ diverges at the critical point with the same exponent $\alpha$ as the specific heat. However, even in the absence of pressure mixing {\em per se} the phase boundary $\mu_{\sigma}(T)$ is not analytic: rather its {\em third derivative diverges} (in contrast again to lattice-gas models) owing to the odd correction-to-scaling term, since in the case of fluids one has $ 2 < 2-\alpha-\beta+\theta_{5}= \Delta + \theta_{5} <3$. [Note that the amplitude, $A_{\mu}b_{\mu}$, of $|t|^{2-\alpha-\beta+\theta_{5}}$ in (\ref{mu-boundary}) does {\em not} vanish when $j_{2}=0$.] The conclusion that $\mu_{\sigma}^{\prime\prime}(T)$ exhibits a cusp-like behavior near the critical point was originally advanced by Ley-Koo and Green \cite{ley-koo} and enters into the analysis of the effects of impurities on the detection of the Yang-Yang anomaly \cite{kos:ani}: see the discussion below in Sec.\ III.D.

\subsection{Densities and entropies at coexistence}
\label{sec3.2}
The generalized densities along the phase boundary, $\tilde{\rho}_{\sigma}$ and $\tilde{s}_{\sigma}$, can be obtained from (\ref{scaling-eq}) and (\ref{gen-den}) by using the results (\ref{p-boundary}) and (\ref{mu-boundary}) for $\check{p}_{\sigma}$ and $\check{\mu}_{\sigma}$. After some algebra, we find
 \begin{eqnarray}
  \tilde{\rho}_{\sigma}(t) & \approx & \pm QU|\tau t|^{\beta}\left[ W_{-1}^{0} + U_{4\mbox{\scriptsize c}}W_{-1}^{(4)}|\tau t|^{\theta} \right. \nonumber \\
  &  &\left. \pm \: U_{5\mbox{\scriptsize c}}\left\{W_{-1}^{(5)}-(W_{-2}^{0}W_{-0}^{(5)}/W_{-1}^{0})\right\}|\tau t|^{\theta_{5}} + \cdots \right], \nonumber \\  \label{gen-rho-sigma} \\
  \tilde{s}_{\sigma}(t) & \approx & - Q|\tau t|^{1-\alpha}\left[(2-\alpha)W_{-0}^{0} + (2-\alpha+\theta)U_{4\mbox{\scriptsize c}}W_{-0}^{(4)}|\tau t|^{\theta} \right. \nonumber \\
  &  & \left. \pm\: (\Delta+\theta_{5})U_{5\mbox{\scriptsize c}}W_{-0}^{(5)}|\tau t|^{\theta_{5}} + \cdots \right],  \label{gen-s-sigma}
 \end{eqnarray}
where $\pm$ refers to $\tilde{h}\gtrless 0$. Using (\ref{checkrho-eq1}) and (\ref{checks-eq1}) and expanding in powers of $t$, we finally obtain the number density and entropy density on the two sides of the coexistence curve as
 \begin{eqnarray}
  \rho^{\pm}(T) & = & \rho_{\mbox{\scriptsize c}}\mbox{\Large $\{$} 1 + A_{2\beta}|t|^{2\beta} + A_{1-\alpha}|t|^{1-\alpha}+ A_{1}t + \cdots \nonumber \\
  &  &  +\: A_{5}|t|^{\beta+\theta_{5}} + \cdots \nonumber \\
  &  & \pm\: B|t|^{\beta}\left[ 1 + b_{\theta}|t|^{\theta}+ b_{2\beta}|t|^{2\beta} + \cdots \right] \mbox{\Large $\}$},  \label{rho-boundary} \\
 {\cal S}^{\pm}(T) & = & \rho_{\mbox{\scriptsize c}}k_{\mbox{\scriptsize B}} \mbox{\Large $\{$} k_{0} + S_{2\beta}|t|^{2\beta} + S_{1-\alpha}|t|^{1-\alpha}  + S_{1}t + \cdots \nonumber \\
  &  & +\: S_{5}|t|^{\beta+\theta_{5}} + \cdots \nonumber \\
  &  & \pm\: B_{s}|t|^{\beta}\left[1 + d_{\theta}|t|^{\theta} + d_{2\beta}|t|^{2\beta} + \cdots \right] \mbox{\Large $\}$},  \label{s-boundary}
 \end{eqnarray}
where terms varying as $|t|^{\beta +1}$ and $|t|^{1-\alpha+\beta}$ have not been displayed and higher order terms such as $\pm|t|^{\beta +\Delta}$, $|t|^{\beta+2-\alpha}$, etc. will also be present in general. Implicit in these results is the conclusion
 \begin{equation}
  l_{0}=\check{\rho}_{\mbox{\scriptsize c}} \equiv 1 \hspace{0.1in} \mbox{and} \hspace{0.1in} k_{0} = \check{s}_{\mbox{\scriptsize c}} \equiv {\cal S}_{\mbox{\scriptsize c}}/\rho_{\mbox{\scriptsize c}}k_{\mbox{\scriptsize B}}. \label{l_0-k_0}
 \end{equation}
The leading coefficients for the density are then
 \begin{eqnarray}
  B & = & (1-j_{2})QUW_{-1}^{0}|\tau|^{\beta}, \hspace{0.1in} b_{\theta} = U_{4\mbox{\scriptsize c}}W_{-1}^{(4)}|\tau|^{\theta}/W_{-1}^{0}, \nonumber \\ \label{rho-coeff-a} \\
  b_{2\beta} & = & j_{2}^{2}B^{2}/(1-j_{2})^{2}, \hspace{0.1in} A_{2\beta} = -\:j_{2}B^{2}/(1-j_{2}),\nonumber \\ \label{rho-coeff-b} \\
  A_{1-\alpha} & = & (2-\alpha)(l_{1}+j_{1})QW_{-0}^{0}|\tau|^{1-\alpha}, \label{rho-coeff} \\
  A_{1} & = & v_{0}+m_{3}+(2q_{0}+n_{0})\check{\mu}_{\sigma,1} + (n_{0}+2m_{0})\check{p}_{\sigma,1}, \nonumber \\ \label{rho-ceoff-c}
 \end{eqnarray}
while the coefficients for the entropy satisfy
 \begin{eqnarray}
  B_{s}/B & = & -\: (k_{1}+j_{2}k_{0})/(1-j_{2}), \hspace{0.1in} d_{\theta} = b_{\theta}, \label{s-coeff-a} \\
  d_{2\beta} & = & j_{2}^{2} B_{s}^{2}/(k_{1}+j_{2}k_{0})^{2}, \hspace{0.1in} S_{2\beta} = -j_{2}B_{s}^{2}/(k_{1}+j_{2}k_{0}), \nonumber \\  \label{s-coeff-b} \\
  S_{1-\alpha} & = & (2-\alpha)(j_{1}k_{0}-1)QW_{-0}^{0}|\tau|^{1-\alpha}, \hspace{0.1in} S_{5} = B_{s} A_{5}/B, \nonumber \\  \label{s-coeff}
 \end{eqnarray}
and $A_{5}$ and $S_{5}$ follow from {\bf K}(3.83, 3.84).

From (\ref{rho-boundary}) the coexistence curve diameter, $\bar{\rho}(T) = \frac{1}{2}[\rho_{\mbox{\scriptsize liq}}(T) + \rho_{\mbox{\scriptsize vap}}(T)]$, and the width, $2\Delta\rho (T)= \rho_{\mbox{\scriptsize liq}}(T) - \rho_{\mbox{\scriptsize vap}}(T)$ of the coexistence curve, follows immediately. As anticipated, we see that the diameter contains a $|t|^{2\beta}$ term that is proportional to the pressure-mixing coefficient $j_{2}$; and, since $2\beta < 1-\alpha$ for typical fluids, this actually dominates the previously anticipated $|t|^{1-\alpha}$ term \cite{mer}. Likewise, one may read off the entropy diameter, $\bar{\cal S}(T)$, and entropy jump, $\Delta{\cal S}(T)$, from (\ref{s-boundary}). Again one observes that the dominating $|t|^{2\beta}$ term in $\bar{\cal S}(T)$ is proportional to $j_{2}$.

The total entropy in the two-phase region is given by
 \begin{equation}
  S_{\sigma}^{\mbox{\scriptsize tot}}(T) = V\frac{dp_{\sigma}}{dT} - N \frac{d\mu_{\sigma}}{dT}.  \label{tot-entropy}
 \end{equation}
On the critical isochore, $\rho=\rho_{\mbox{\scriptsize c}}$, this yields the entropy density
 \begin{equation}
  {\cal S}_{\sigma}(T;\rho_{\mbox{\scriptsize c}}) = {\cal S}_{\mbox{\scriptsize c}} + \rho_{\mbox{\scriptsize c}}k_{\mbox{\scriptsize B}}(1-j_{2})A_{p}|t|^{1-\alpha} + {\cal O}(t)  \label{tot-s}
 \end{equation}
from which, as was to be anticipated, the $|t|^{2\beta}$ terms have cancelled. [Recall that $A_{p}$ is defined in (3.13).]

\begin{figure}[h]
\vspace{-0.8in}
\centerline{\epsfig{figure=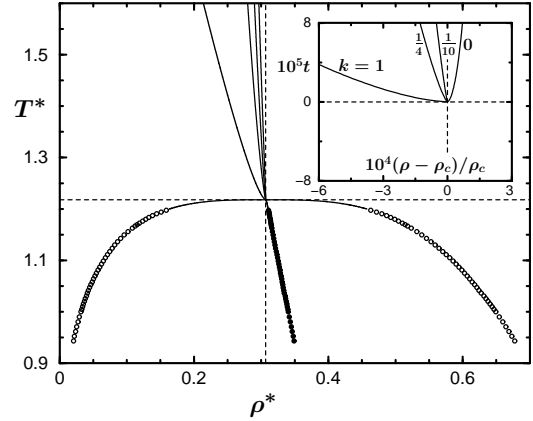,width=3.2in,angle=0}}
\vspace{-0.8in}

\caption {Coexistence data (open circles, with closed circles for the diameter) for the hard-core square-well fluid obtained by Orkoulas {\em et al.} [6]. The solid lines connecting the data points represent an Ising-type fit: see Eq.\ (3.20). The vertical and horizontal dashed-lines locate the critical isochore, $\rho^{\ast}=\rhoc^{\ast} \simeq 0.3067$, and the critical isotherm, $T^{\ast}=\Tc^{\ast}\simeq 1.2179$ [6]. The curves above criticality and in the inset depict estimates for the $k$-susceptibility loci for the values of $k$ indicated.}
\label{fig1}
\end{figure}

\begin{figure}[h]
\vspace{-0.8in}
\centerline{\epsfig{figure=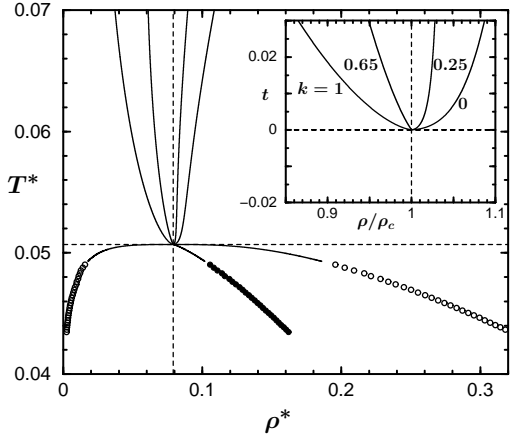,width=3.2in,angle=0}}
\vspace{-0.8in}
 \caption {Plots of coexistence curves, $k$-susceptibility loci, etc., as in Fig.\ \ref{fig1}, based on simulations for the restricted primitive model by Luijten {\em et al.} [25]. The critical parameters adopted are $\Tc^{\ast} \simeq 0.05069$ and $\rhoc^{\ast} \simeq 0.079$ corresponding to a $\zeta=5$ fine-discretization level [25].}
\label{fig2}
\end{figure}

In Fig.\ \ref{fig1} grand canonical simulation data for the coexistence curve of the hard-core square-well fluid are presented. Adopting the critical point estimates, $\Tc \simeq 1.2179$ and $\rhoc \simeq 0.3067$ \cite{ork:fis:pan}, and the Ising values for the critical exponents yields the estimates $B=1.2026_{4}$, $A_{2\beta}=-0.0007_{3}$, $A_{1-\alpha}=0.189_{7}$, $A_{1}=-0.0691_{4}$, $b_{\theta}=-0.257_{6}$, and $b_{2\beta}=-0.085_{2}$ for the amplitudes in (\ref{rho-boundary}) \cite{ork:fis:pan}.
 This fit is shown in Fig.\ \ref{fig1} as a solid line that connects the coexisting density estimates (circles) to the critical point. Similarly, coexistence simulation data (in a $\zeta=5$ fine discretization level) for the restricted primitive model electrolyte are presented in Fig.\ \ref{fig2}.
 The solid line --- the Ising fit to the coexistence data --- is drawn by adopting the critical point values, $\Tc \simeq 0.05069$, $\rhoc \simeq 0.079$ \cite{luijten2} and the amplitude estimates, $B = 0.274$, $A_{2\beta} = 0.0165$, $A_{1-\alpha} = 0.919$, $A_{1} = 0.586$, $b_{\theta} = 1.464$, and $b_{2\beta}=2.254$. However, one must note that the specific numbers attached to these particular amplitude estimates {\em cannot} have a very significant meaning, unless the higher order corrections are considered more carefully than is practicable with the data currently available.

\subsection{Mixing coefficients}
\label{sec3.3}
It is clearly of interest to express the linear mixing coefficients entering the scaling fields (\ref{tilde-g})-(\ref{tilde-h}) in terms of various thermodynamic quantities which might, at least in principle, be measured in experiments or simulations. As regards the field $\tilde{p}$, we have already noted in (3.22) the simple expressions for $k_{0}$ and $l_{0}$.

The pressure-mixing coefficient, $j_{2}$, can be obtained via (3.16) from the singular amplitudes $A_{\mu}$ in $\mu_{\sigma}(T)$ and $A_{p}$ in $p_{\sigma}(T)$ simply as
 \begin{equation}
  j_{2} = A_{\mu}/A_{p}.  \label{j2}
 \end{equation}
Then, from the observable limiting derivatives $\mu_{\sigma\mbox{\scriptsize c}}^{\prime} \equiv (d\mu_{\sigma}/dT)_{\mbox{\scriptsize c}}$ and $p_{\sigma\mbox{\scriptsize c}}^{\prime} \equiv (dp_{\sigma}/dT)_{\mbox{\scriptsize c}}$, which correspond to $\check{\mu}_{\sigma,1}$ and $\check{p}_{\sigma,1}$, we may obtain
 \begin{equation}
  k_{1} = (\rhoc \mu_{\sigma\mbox{\scriptsize c}}^{\prime} -j_{2}p_{\sigma\mbox{\scriptsize c}}^{\prime})/\rhoc k_{\mbox{\scriptsize B}} = [\rhoc\mu_{\sigma\mbox{\scriptsize c}}^{\prime} - p_{\sigma\mbox{\scriptsize c}}^{\prime}(A_{\mu}/A_{p})]/\rhoc k_{\mbox{\scriptsize B}}.  \label{k1}
 \end{equation}

The remaining two linear mixing coefficients, $j_{1}$ and $l_{1}$, can be obtained by using the amplitudes, $A_{1-\alpha}$ and $S_{1-\alpha}$ in (\ref{rho-coeff}) and (\ref{s-coeff}) which describe the $|t|^{1-\alpha}$ singularity in the density and entropy diameters. Taking a ratio yields
 \begin{equation}
  \frac{l_{1} + j_{1}}{j_{1}k_{0} - 1} = \frac{A_{1-\alpha}}{S_{1-\alpha}},  \label{Ad-As}
 \end{equation}
where we have used $l_{0} = 1$, while $k_{0}$ is given in (3.22). On the other hand, from the ratio of $A_{1-\alpha}$ to $A_{p}$ we obtain via (3.13) and (3.25) the distinct relation
 \begin{equation}
  \frac{l_{1} + j_{1}}{|\tau|} = \frac{A_{1-\alpha}}{(2-\alpha)(1-j_{2})A_{p}},  \label{As-A1}
 \end{equation}
where $\tau$ is given in (\ref{tau1}). Since $\tau$ is linear in the mixing coefficients $j_{1}$ and $l_{1}$ one can, in principle, solve these two equations for $j_{1}$ and $l_{1}$. In practice, however, lack of precision in measuring the amplitudes $A_{p}$, $A_{1-\alpha}$ and $S_{1-\alpha}$ is likely to produce large uncertainties.

\subsection{Yang-Yang anomaly}
\label{sec3.4}

The Yang-Yang relation (\ref{yang-yang}) in the two-phase region $(T<\Tc)$ can be usefully rewritten as \cite{fis:ork}
 \begin{equation}
  C_{V}(T,\rho) \equiv C_{V}^{\mbox{\scriptsize tot}} = (v/v_{\mbox{\scriptsize c}})\tilde{C}_{p}(T) + \tilde{C}_{\mu}(T),  \label{yang-2}
 \end{equation}
where $v = V/N = 1/\rho$ and
 \begin{equation}
  \tilde{C}_{p}(T) \equiv v_{\mbox{\scriptsize c}}T(d^{2}p_{\sigma}/dT^{2}), \hspace{0.1in}  \tilde{C}_{\mu}(T) \equiv -T(d^{2}\mu_{\sigma}/dT^{2}).
 \end{equation}
The results (3.11) and (3.15) then yield
 \begin{eqnarray}
  \tilde{C}_{p}(T) & = & \tilde{A}_{p}|t|^{-\alpha} + \tilde{B}_{p} + \tilde{a}_{p}|t|^{\theta-\alpha} + \cdots \nonumber \\
  &  & +\: \tilde{b}_{1}|t|^{\theta_{5}-\alpha-\beta}  + \tilde{b}_{p}|t|^{\theta_{5}-\alpha-\beta+\theta} + \cdots,  \label{tilde_Cp} \\
  \tilde{C}_{\mu}(T) & = & \tilde{A}_{\mu}|t|^{-\alpha} + \tilde{B}_{\mu} + \tilde{a}_{\mu}|t|^{\theta-\alpha} + \cdots \nonumber \\
  &  & +\: \tilde{c}_{1}|t|^{\theta_{5}-\alpha-\beta} + \tilde{b}_{\mu}|t|^{\theta_{5}-\alpha-\beta-\theta} + \cdots,  \label{tilde-Cmu}
 \end{eqnarray}
where the various coefficients follow straightforwardly from (3.12)-(3.14), (3.16) and (3.17): see also {\bf K}(3.103). Following Ref.\ [2] it is reasonable to define the strength of the Yang-Yang anomaly via
 \begin{equation}
  {\cal R}_{\mu} \equiv \lim_{t\rightarrow 0-} \frac{\tilde{C}_{\mu}(T)}{\tilde{C}_{p}+\tilde{C}_{\mu}} = \frac{\tilde{A}_{\mu}}{\tilde{A}_{p}+\tilde{A}_{\mu}}. \label{Yang-anomaly}
 \end{equation}
By (3.32) this leads immediately to
 \begin{equation}
  {\cal R}_{\mu} = -j_{2}/(1-j_{2}).  \label{yang-anomaly2}
 \end{equation}
Note that ${\cal R}_{\mu}$ depends only on $j_{2}$ (not on $j_{1}$). Fisher and Orkoulas \cite{fis:ork} estimated ${\cal R}_{\mu}$ for propane from experimental data on the two-phase heat capacity and obtained ${\cal R}_{\mu} \simeq 0.56$: this suggests $j_{2} \simeq -1.27$. They also analyzed the heat capacity data for CO$_{2}$ and estimated ${\cal R}_{\mu} \simeq -0.4 \; (\pm 0.3)$ which implies that $j_{2}$ should be positive but small (less than 1). On the other hand, simulations of the hard-core square-well fluid \cite{ork:fis:pan} indicate that ${\cal R}_{\mu}$ is small but negative, close to zero: correspondingly, $j_{2}$ should be small but positive. 

\section{Special Critical Loci}
\label{sec4}
In this section we use the scaling formulation to obtain asymptotic expressions for various interesting critical loci that lie in the one-phase regions of the phase diagram. For convenience, we introduce a superscript index $\iota$ defined so that: $\,\iota= \mbox{\bf i}$ for the locus $\mu=\mu_{\mbox{\scriptsize c}}$, say, the critical {\em isokyme}; $\,\iota=\mbox{\bf ii}$ for the critical isobar, $p=p_{\mbox{\scriptsize c}}$; $\,\iota=\mbox{\bf iii}$ for the critical isotherm, $T=\Tc$; and $\,\iota=\mbox{\bf iv}$ for the critical isochore, $\rho=\rhoc$.

\subsection{Critical isokyme, {\boldmath $\mu = \mu_{\mbox{\scriptsize c}}$}}
\label{sec4.1}
On the critical isokyme $\mu = \mu_{\mbox{\scriptsize c}}$ (or $\check{\mu} = 0$) the scaling fields (\ref{tilde-g})-(\ref{tilde-h}) reduce to
 \begin{eqnarray}
  \tilde{p} & = & \check{p} - k_{0}t - m_{0}\check{p}^{2} - r_{0}t^{2} - n_{3}\check{p}t + \cdots,  \label{muc-tildep}  \\
  \tilde{h} & = & -\: j_{2}\check{p} - k_{1}t - m_{2}\check{p}^{2} - r_{2}t^{2} - n_{5}\check{p}t + \cdots,  \label{muc-tildemu} \\
  \tilde{t} & = & t - j_{1}\check{p} - m_{1}\check{p}^{2} - r_{1}t^{2} - n_{4}\check{p}t + \cdots.  \label{muc-tildet}
 \end{eqnarray}
Notice that the scaling variable $y = U\tilde{h}/|\tilde{t}|^{\Delta} \sim -(j_{2}\check{p} + k_{1}t)/|t-j_{1}\check{p}|^{\Delta}$ now diverges, in general, when the critical point is approached ({\em i.e.,} when $\check{p}$, $t\rightarrow 0$) since $\Delta >1$ for fluids. Hence, we need the expansions of the scaling functions $W_{\pm}^{\mbox{\scriptsize \boldmath $\kappa$}}(y)$ for $y\rightarrow\infty$. On using (2.3), (2.4) and (\ref{Wpm-highexpan}), we obtain
 \begin{eqnarray}
  \check{p} & = & k_{0}t + m_{0}\check{p}^{2} + r_{0}t^{2} + n_{3}\check{p}t + \cdots \nonumber \\
  & & +\: QW_{\infty}^{0}(U|\tilde{h}|)^{(2-\alpha)/\Delta}\left[ 1 + w_{1}^{0}\tilde{t}(U|\tilde{h}|)^{-1/\Delta} + \cdots \right] \nonumber \\
  &  & +\: QW_{\infty}^{(4)}U_{4\mbox{\scriptsize c}}(U|\tilde{h}|)^{(2-\alpha+\theta)/\Delta}\left[ 1 + \cdots \right],  \nonumber \\
  &  & +\: \tilde{\sigma}_{h}QW_{\infty}^{(5)}U_{5\mbox{\scriptsize c}}(U|\tilde{h}|)^{(2-\alpha+\theta_{5})/\Delta}\left[ 1 + \cdots \right],  \label{muc-p1}
 \end{eqnarray}
where $\tilde{\sigma}_{h} = \mbox{sgn}(\tilde{h})$. Now $\tilde{t}$ and $\tilde{h}$ can be expanded in terms of $\check{p}$ and $t$ using (\ref{muc-tildemu}) and (\ref{muc-tildet}) and the resulting equation may then be solved iteratively for $\check{p}$ as a function of $t$. After some algebra, this yields the critical isokyme in the $(p,T)$ plane as
 \begin{eqnarray}
  \check{p}^{\mbox{\scriptsize \bf i}}(t) & \equiv & [p^{\mbox{\scriptsize \bf i}}(T) - p_{\mbox{\scriptsize c}}]/\rhoc k_{\mbox{\scriptsize B}}\Tc \nonumber \\
  & = & \check{p}_{1}^{\mbox{\scriptsize \bf i}}t + \check{p}_{2}^{\mbox{\scriptsize \bf i}}t^{2} + A_{p}^{\mbox{\scriptsize \bf i}}|t|^{(2-\alpha)/\Delta} \mbox{\Large $[$}1 \pm b_{1}^{\mbox{\scriptsize \bf i}}|t|^{\beta/\Delta} \nonumber \\
  &  & \pm\: b_{2}^{\mbox{\scriptsize \bf i}}|t|^{(\Delta -1)/\Delta}  + \cdots +  b_{3}^{\mbox{\scriptsize \bf i}}|t|^{\theta/\Delta} \pm b_{4}^{\mbox{\scriptsize \bf i}}|t|^{(\beta+\theta)/\Delta} + \nonumber \\
  &  & \cdots  +\ b_{5}^{\mbox{\scriptsize \bf i}}|t|^{\theta_{5}/\Delta} + \: b_{6}^{\mbox{\scriptsize \bf i}}|t|^{(\beta+\theta_{5})/\Delta} + \cdots \mbox{\Large $]$} + \cdots,  \label{muc-p2}
 \end{eqnarray}
where $\pm$ refers to $t\gtrless 0$ while
 \begin{eqnarray}
  \check{p}_{1}^{\mbox{\scriptsize \bf i}} & = & k_{0}, \hspace{0.2in} \check{p}_{2}^{\mbox{\scriptsize \bf i}} = r_{0} + k_{0}^{2}m_{0} + k_{0}n_{3}, \nonumber \\
  A_{p}^{\mbox{\scriptsize \bf i}} & = & QW_{\infty}^{0}(U|k_{1} + j_{2}k_{0}|)^{(2-\alpha)/\Delta}, \label{muc-p3}
 \end{eqnarray}
and the amplitudes $b_{1}^{\mbox{\scriptsize \bf i}}$, $\cdots$ of the correction terms are given in {\bf K}(3.111).

Note that the leading singular exponent is $(2-\alpha)/\Delta = 1 + \delta^{-1}$, where $\delta = \Delta/\beta$ is the standard critical exponent characterizing the critical isotherm. In the case of the ($d=3$)-dimensional Ising universality class, $\delta \simeq 4.8$ so that $1 + \delta^{-1} \simeq 1.21$. This implies that the curvature of $\check{p}^{\mbox{\scriptsize \bf i}}(t)$, the pressure on the critical isokyme, diverges as $T\rightarrow T_{\mbox{\scriptsize c}}$. Also by convexity --- see Sec.\ II.A --- the leading singular amplitude $A_{p}^{\mbox{\scriptsize \bf i}}$ is positive regardless of the sign of $t$. On the other hand, it transpires that the amplitude $b_{5}^{\mbox{\scriptsize \bf i}}$ of the leading odd correction contains the signum factor $\tilde{\sigma}_{h}$ so that its sign depends on $\tilde{h}$. Later we will see that the sign of $\tilde{h}$ can be determined from the mixing coefficient $k_{1}$.

To obtain the critical isokyme in the $(\rho,T)$ plane, we first substitute (\ref{muc-p2}) into (\ref{muc-tildemu}) and (\ref{muc-tildet}) and express $\tilde{h}$ and $\tilde{t}$ as functions of $t$. Using these results in the definitions (\ref{gen-den}) then yields $\tilde{\rho}$ and $\tilde{s}$ in terms of $t$. Finally, from (\ref{checkrho-eq1}), we obtain the density on the critical isokyme, $\mu=\mu_{\mbox{\scriptsize c}}$, as
 \begin{eqnarray}
 \rho^{\mbox{\scriptsize \bf i}}(T) & = & \rho_{\mbox{\scriptsize c}} \mbox{\Large $\{$} 1 \pm B^{\mbox{\scriptsize \bf i}}|t|^{\beta/\Delta} + B_{d}^{\mbox{\scriptsize \bf i}}|t|^{2\beta/\Delta} + A_{d}^{\mbox{\scriptsize \bf i}}|t|^{(1-\alpha)/\Delta} + \nonumber \\
   &  & \cdots \pm B_{4}^{\mbox{\scriptsize \bf i}}|t|^{(\beta+\theta)/\Delta} + \cdots + B_{5}^{\mbox{\scriptsize \bf i}}|t|^{(\beta+\theta_{5})/\Delta} + \cdots \mbox{\Large $\}$},\nonumber \\  \label{muc-rho}
 \end{eqnarray}
where $\pm$ refers to $\tilde{h} \gtrless 0$, while the leading coefficients are
 \begin{eqnarray}
  B^{\mbox{\scriptsize \bf i}} & = & (1+\delta^{-1})(1-j_{2})A_{p}^{\mbox{\scriptsize \bf i}}/|k_{1}+j_{2}k_{0}|,  \nonumber \\
  B_{d}^{\mbox{\scriptsize \bf i}} & = & -j_{2}(2-\alpha+\beta)(B^{\mbox{\scriptsize \bf i}})^{2}/(2-\alpha)(1-j_{2}), \label{muc-rho-coeff}
 \end{eqnarray}
and the further coefficients are presented in {\bf K}(3.113).

Note that the leading singular exponent is $\beta/\Delta = \delta^{-1}$ which is less than $\beta$ so implying that the the critical isokyme in the $(\rho,T)$ plane is significantly flatter than the coexistence curve. Clearly, therefore, the critical isokyme {\em below} $\Tc$ lies outside the coexistence boundary, {\em i.e.,} entirely in the one-phase region as is to be expected.

\subsection{Critical isobar}
\label{sec4.2}

At $p = p_{\mbox{\scriptsize c}}$ (or $\check{p}=0$), the scaling fields (\ref{tilde-g})-(\ref{tilde-h}) with (3.22) reduce to
 \begin{eqnarray}
  \tilde{p} & = & -\: k_{0}t - \check{\mu} - q_{0}\check{\mu}^{2} - r_{0}t^{2} - v_{0}\check{\mu}t + \cdots,  \label{pc-tildeg} \\
  \tilde{h} & = & \check{\mu} - k_{1}t - q_{2}\check{\mu}^{2} - r_{2}t^{2} - v_{2}\check{\mu}t + \cdots,  \label{pc-tildeh} \\
  \tilde{t} & = & t - l_{1}\check{\mu} - q_{1}\check{\mu}^{2} - r_{1}t^{2} - v_{1}\check{\mu}t + \cdots.  \label{pc-tildet}
 \end{eqnarray}
As the critical point is approached along a general locus ({\em i.e.,} $t,\,\check{\mu} \rightarrow 0$), the scaling variable $y \propto \tilde{h}/|\tilde{t}|^{\Delta} \sim (\check{\mu}-k_{1}t)/|t-l_{1}\check{\mu}|^{\Delta}$ again diverges. Using the large-$y$ expansion (\ref{Wpm-highexpan}) for $\tilde{p}$ and rearranging (\ref{pc-tildeg}) yields
 \begin{eqnarray}
  \check{\mu} & = & -\: k_{0}t-q_{0}\check{\mu}^{2}-r_{0}t^{2}-v_{0}\check{\mu}t + \cdots \nonumber \\
  &  & - \: QW_{\infty}^{0}(U|\tilde{h}|)^{(2-\alpha)/\Delta}\left[ 1 + w_{1}^{0}\tilde{t}(U|\tilde{h}|)^{-1/\Delta} + \cdots\right] \nonumber  \\
  &  & - \: QW_{\infty}^{(4)}U_{4\mbox{\scriptsize c}}(U|\tilde{h}|)^{(2-\alpha+\theta)/\Delta}[1 + \cdots ] \nonumber \\
  &  & - \: QW_{\infty}^{(5)}U_{5\mbox{\scriptsize c}}(U|\tilde{h}|)^{(2-\alpha+\theta_{5})/\Delta}[ 1+ \cdots]. \label{pc-mu1}
 \end{eqnarray}
Solving this equation iteratively for $\check{\mu}$ in terms of $t$ with the aid of (\ref{pc-tildeh}) and (\ref{pc-tildet}), expresses the critical isobar in the $(\mu,T)$ plane as
 \begin{eqnarray}
  \check{\mu}^{\mbox{\scriptsize \bf ii}}(t) & = & [\mu^{\mbox{\scriptsize \bf ii}}(T)-\mu_{\mbox{\scriptsize c}}]/k_{\mbox{\scriptsize B}}\Tc \nonumber \\
  & = & -\: k_{0}t + \check{\mu}_{2}^{\mbox{\scriptsize \bf ii}}t^{2} + \cdots + A_{\mu}^{\mbox{\scriptsize \bf ii}}|t|^{(2-\alpha)/\Delta}\mbox{\Large $[$} 1 \pm b_{1}^{\mbox{\scriptsize \bf ii}}|t|^{\beta/\Delta} \nonumber \\
  &  &  \pm\:  b_{2}^{\mbox{\scriptsize \bf ii}}|t|^{(\Delta-1)/\Delta} + \cdots  + b_{3}^{\mbox{\scriptsize \bf ii}}|t|^{\theta/\Delta} \pm b_{4}^{\mbox{\scriptsize \bf ii}}|t|^{(\beta+\theta)/\Delta} \nonumber \\
  &  & + \cdots  +  b_{5}^{\mbox{\scriptsize \bf ii}}|t|^{\theta_{5}/\Delta} + b_{6}^{\mbox{\scriptsize \bf ii}}|t|^{(\beta+\theta_{5})/\Delta} + \cdots \mbox{\Large $]$},  \label{pc-mu2}
 \end{eqnarray}
where $\pm$ again refers to $t\gtrless 0$, while the leading coefficients are
 \begin{eqnarray}
  \check{\mu}_{2}^{\mbox{\scriptsize \bf ii}} & = &  -\: r_{0} + k_{0}v_{0} - k_{0}^{2}q_{0},  \nonumber \\
  A_{\mu}^{\mbox{\scriptsize \bf ii}} & = & -\: QW_{\infty}^{0}(U|k_{1}+k_{0}|)^{(2-\alpha)/\Delta},\label{pc-mu3} \\
  b_{1}^{\mbox{\scriptsize \bf ii}} & = &  - (2-\alpha)A_{\mu}^{\mbox{\scriptsize \bf ii}}/\Delta (k_{1}+k_{0}), \nonumber
 \end{eqnarray}
and the further correction amplitudes are given in {\bf K}(3.179).

Note that the curvature of $\check{\mu}^{\mbox{\scriptsize \bf ii}}(t)$ diverges at the critical point with the same exponent as does the critical isokyme in the $(p,T)$ plane: see (\ref{muc-p2}). The critical isobar also has the same sign of curvature {\em above} and {\em below} $\Tc$ in the $(\mu,T)$ plane. By thermodynamic convexity, the leading singular amplitude $A_{\mu}^{\mbox{\scriptsize \bf ii}}$ is negative. The amplitude $b_{5}^{\mbox{\scriptsize \bf ii}}$ of the leading odd correction term again changes its sign depending on $\tilde{h}$.
 
In the $(\rho,T)$ plane the critical isobar can be obtained via the same route used above for the critical isokyme. The result is 
 \begin{eqnarray}
  \rho^{\mbox{\scriptsize \bf ii}}(T) &  = & \rhoc \mbox{\Large $\{$} 1 \pm B^{\mbox{\scriptsize \bf ii}}|t|^{\beta/\Delta} + B_{d}^{\mbox{\scriptsize \bf ii}}|t|^{2\beta/\Delta} + A_{d}^{\mbox{\scriptsize \bf ii}}|t|^{(1-\alpha)/\Delta} + \nonumber \\
  &  & \cdots \pm B_{4}^{\mbox{\scriptsize \bf ii}}|t|^{(\beta+\theta)/\Delta} + \cdots + B_{5}^{\mbox{\scriptsize \bf ii}}|t|^{(\beta+\theta_{5})/\Delta} + \cdots \mbox{\Large $\}$}, \nonumber \\ \label{pc-rho1}
 \end{eqnarray}
where $\pm$ here refers to $\mu\gtrless \mu_{\mbox{\scriptsize c}}$ while the leading coefficients are
 \begin{eqnarray}
  B^{\mbox{\scriptsize \bf ii}} & = & -(1+\delta^{-1})(1-j_{2})A_{\mu}^{\mbox{\scriptsize \bf ii}}/|k_{1}+k_{0}|, \nonumber  \\
  B_{d}^{\mbox{\scriptsize \bf ii}} & = & -[(2-\alpha)j_{2}+\beta] (B^{\mbox{\scriptsize \bf ii}})^{2}/(2-\alpha)(1-j_{2}), \label{pc-rho2}
 \end{eqnarray}
where the further coefficients are to be found in {\bf K}(3.121).

Notice that the leading singular behavior matches that of the density on the critical isokyme as given in (\ref{muc-rho}). For small $j_{2}\, (<1)$, the leading amplitude $B^{\mbox{\scriptsize \bf ii}}$ must, by convexity, again be positive. The ratio between $B^{\mbox{\scriptsize \bf i}}$ in (4.7) and $B^{\mbox{\scriptsize \bf ii}}$ is simply
 \begin{equation}
  B^{\mbox{\scriptsize \bf i}}/B^{\mbox{\scriptsize \bf ii}} = \left| (k_{1}+j_{2}k_{0})/(k_{1}+k_{0})\right|^{\beta/\Delta}.  \label{amplitude-ratio}
 \end{equation}

\subsection{Critical isotherm}
\label{sec4.3}
When $T=\Tc$ so that $t=0$, the scaling fields again reduce to yield, now,
 \begin{equation}
  \tilde{p} = \check{p} - \check{\mu} - m_{0}\check{p}^{2} - q_{0}\check{\mu}^{2} - n_{0}\check{p}\check{\mu} + \cdots,  \label{Tc-tildeg}  \label{Tc-tildet}
 \end{equation}
and similarly for $\tilde{h}$ and $\tilde{t}$. Once more, the scaling variable $y$ diverges, in general, on approach to criticality. Using (4.18) and the large-$y$ expansion (\ref{Wpm-highexpan}) for the scaling functions yields an equation for $\check{p}$ in powers of $\check{\mu}$, $\tilde{h}$, $\tilde{t}$ and $\check{p}$ which, as before, can be solved iteratively with the aid of the reduced expansions for $\tilde{h}$ and $\tilde{t}$ to obtain $\check{p}$ as a function of $\check{\mu}$. The result for the pressure on the critical isotherm is
 \begin{eqnarray}
  \check{p}^{\mbox{\scriptsize \bf iii}}(\check{\mu}) & = & [p^{\mbox{\scriptsize \bf iii}}(\mu) - p_{\mbox{\scriptsize c}}]/\rhoc k_{\mbox{\scriptsize B}}\Tc \nonumber \\
   & = & \check{\mu} + \check{p}_{2}^{\mbox{\scriptsize \bf iii}}\check{\mu}^{2} + \cdots \nonumber \\
  &  & +\: A_{p}^{\mbox{\scriptsize \bf iii}}|\check{\mu}|^{(2-\alpha)/\Delta} \left[ 1 \pm b_{1}^{\mbox{\scriptsize \bf iii}}|\check{\mu}|^{\beta/\Delta} \pm \cdots  \right],  \label{Tc-p2}
 \end{eqnarray}
where $\pm$ refers to $\mu \gtrless \mu_{\mbox{\scriptsize c}}$, while the coefficients are
 \begin{eqnarray}
  \check{p}_{2}^{\mbox{\scriptsize \bf iii}} & = & m_{0} + q_{0} + n_{0}, \nonumber  \\
  A_{p}^{\mbox{\scriptsize \bf iii}} & = & QW_{\infty}^{0}(U|1-j_{2}|)^{(2-\alpha)/\Delta}, \label{Tc-p3} \\
  b_{1}^{\mbox{\scriptsize \bf iii}} & = & -\: j_{2}(2-\alpha) A_{p}^{\mbox{\scriptsize \bf iii}}/\Delta (1-j_{2}),  \nonumber
 \end{eqnarray}
and the spectrum of higher order terms matches that in (4.13): see also {\bf K}(3.127). The leading singular exponent is again $(2-\alpha)/\Delta = 1+\delta^{-1}$, implying that the critical isotherm in the $(p,\mu)$ plane has a divergent curvature at the critical point, and thermodynamic convexity ensures $A_{p}^{\mbox{\scriptsize \bf iii}} > 0$.

To obtain the critical isotherm in the $(\rho,p)$ plane, we first invert (\ref{Tc-p2}) to obtain $\,\check{\mu} \approx \check{p} - A_{p}^{\mbox{\scriptsize \bf iii}}|\check{p}|^{(2-\alpha)/\Delta} \,$ and then use this to express the generalized densities, $\tilde{\rho}$ and $\tilde{s}$, as functions of $\check{p}$. Using (\ref{checkrho-eq1}) finally yields the critical isotherm in the form
 \begin{eqnarray}
  \rho^{\mbox{\scriptsize \bf iii}}(p)  & = & \rhoc \left\{ 1 \pm B^{\mbox{\scriptsize \bf iii}}|\check{p}|^{\beta/\Delta} + B_{d}^{\mbox{\scriptsize \bf iii}}|\check{p}|^{2\beta/\Delta} + A_{d}^{\mbox{\scriptsize \bf iii}}|\check{p}|^{(1-\alpha)/\Delta} \right.  \nonumber  \\
   &  &  \hspace{0.35in} + \cdots  \pm B_{4}^{\mbox{\scriptsize \bf iii}}|\check{p}|^{(\beta+\theta)/\Delta} + \cdots \nonumber \\
   &  &  \left. \hspace{0.35in} +\: B_{5}^{\mbox{\scriptsize \bf iii}}|\check{p}|^{(\beta+\theta_{5})/\Delta} + \cdots \right\},  \label{Tc-rho}
 \end{eqnarray}
where $\pm$ again refers to $\mu \gtrless \mu_{\mbox{\scriptsize c}}$ while
 \begin{equation}
  B^{\mbox{\scriptsize \bf iii}} = (1+\delta^{-1})|(1-j_{1})/(1-j_{2})|^{\beta/\Delta} A_{p}^{\mbox{\scriptsize \bf iii}}, \label{Tc-rho2}
 \end{equation}
and $B_{d}^{\mbox{\scriptsize \bf iii}}\propto - (B^{\mbox{\scriptsize \bf iii}})^{2}$: see {\bf K}(3.131). As expected for the critical isotherm the leading exponent is again $\beta/\Delta = \delta^{-1}$. Assuming that the pressure-mixing coefficient $j_{2}$ is small $(<1)$, we find $B^{\mbox{\scriptsize \bf iii}} >0$ owing to convexity. Now when $j_{2}\simeq0$, the scaling field $\tilde{h}$ can be approximated by $\check{p}$ which implies that the critical isotherm in the $(\rho,p)$ plane approaches the critical point from higher density above $p_{\mbox{\scriptsize c}}$, while from lower density below $p_{\mbox{\scriptsize c}}$: this accords with the observed standard behavior.

\subsection{Critical isochore}
\label{sec4.4}
Below $\Tc$ in normal fluids the critical isochore in the $(\mu,T)$ and $(p,T)$ planes coincides with the phase boundary, $\mu_{\sigma}(T)$ and $p_{\sigma}(T)$ (but see \cite{fis:fel} for exceptions in certain models); however, the behavior above $\Tc$ is of general interest. When $\rho=\rhoc$, the result (\ref{checkrho-eq1}) for the density leads to
 \begin{equation}
 \begin{array}{ccl}
  0 & = & (1-j_{2})QU|\tilde{t}|^{\beta}\left[W_{+}^{0\,\prime}(y) +U_{4\mbox{\scriptsize c}}|\tilde{t}|^{\theta}W_{+}^{(4)\,\prime}(y) + \cdots\right]  \\
  &  &+\: j_{2}(j_{2}-1)Q^{2}U^{2}|\tilde{t}|^{2\beta}\mbox{\LARGE $[$}W_{+}^{0\,\prime}(y)+ \cdots \mbox{\LARGE $]$}^{2} \\
  &  & +\: (l_{1} + j_{1})Q|\tilde{t}|^{1-\alpha}\mbox{\LARGE $[$} \Delta y W_{+}^{0\,\prime}(y) - (2-\alpha)W_{+}^{0}(y) \mbox{\LARGE $]$} \nonumber \\
  &  & + \cdots,  \label{rhoc-1} \\
 \end{array}
 \end{equation}
where the primes denote differentiation with respect to $y$. We need to solve this equation for $y$ as a function of $\tilde{t}$; but on the critical isochore we expect $y \rightarrow 0$ when $\tilde{t}\rightarrow 0$. Hence we should now use the {\em small}-$y$ expansions (2.8) for the scaling functions $W_{+}^{0}(y)$, $W_{+}^{(4)}(y)$, etc. The resulting equation in powers of $y$ may be solved iteratively to obtain
 \begin{equation}
  y = Y_{1}|\tilde{t}|^{1-\alpha-\beta} \left[ 1 + y_{2}|\tilde{t}|^{\theta} + \cdots \right],  \label{rhoc-eq3}
 \end{equation}
with coefficients
 \begin{equation}
  Y_{1} = \frac{(2-\alpha)(l_{1}+j_{1})W_{+0}^{0}}{2(1-j_{2})W_{+2}^{0}U}, \hspace{0.1in} y_{2} = -\frac{U_{4\mbox{\scriptsize c}}W_{+2}^{(4)}}{W_{+2}^{0}}.  \label{rhoc-eq4}
 \end{equation}

The scaling fields $\tilde{p}$ and $\tilde{h}$ along the critical isochore can now be found as follows: first, on combining (\ref{scaling-eq}), (2.4) and (2.8) with (\ref{rhoc-eq3}), we find
 \begin{eqnarray}
  \tilde{p} & = & Q|\tilde{t}|^{(2-\alpha)}\left[ W_{+0}^{0} + W_{+2}^{0}Y_{1}^{2}|\tilde{t}|^{2-2\alpha-2\beta} + \cdots \right. \nonumber \\
  &  &  +\: U_{4\mbox{\scriptsize c}}W_{+0}^{(4)}|\tilde{t}|^{\theta} + \cdots + U_{5\mbox{\scriptsize c}}W_{+1}^{(5)}Y_{1}|\tilde{t}|^{1-\alpha-\beta+\theta_{5}} \nonumber \\
  &  & + \cdots ];  \label{rhoc-tildeg}
 \end{eqnarray}
then, from the definition (2.1) of the scaling variable $y$ we get
 \begin{equation}
  \tilde{h} = (Y_{1}/U)|\tilde{t}|^{1-\alpha+\gamma}\left[ 1 + y_{2}|\tilde{t}|^{\theta} + \cdots \right]; \label{rhoc-tildeh}
 \end{equation}
finally, by expressing the scaling fields in terms of $\check{p}$, $\check{\mu}$ and $t$ via (\ref{tilde-g})-(\ref{tilde-h}), we can solve these two equations iteratively for $\check{p}$ and $\check{\mu}$ as functions of $t$. After some algebra, this yields the critical isochore in the $(p,T)$ plane as
 \begin{eqnarray}
  \check{p}^{\mbox{\scriptsize \bf iv}}(t) & = & [p^{\mbox{\scriptsize \bf iv}}(T)-p_{\mbox{\scriptsize c}}]/\rhoc k_{\mbox{\scriptsize B}}\Tc  \nonumber \\
   & = & \check{p}_{\sigma,1}t + \check{p}_{\sigma,2}t^{2} + \cdots + A_{p}^{\mbox{\scriptsize \bf iv}}|t|^{(2-\alpha)}\mbox{\Large $[$} 1 + a_{1p}^{\mbox{\scriptsize \bf iv}}|t|^{\theta} \nonumber \\
  &  & +\: a_{2p}^{\mbox{\scriptsize \bf iv}}|t|^{\gamma-\alpha} + \cdots  + a_{3p}^{\mbox{\scriptsize \bf iv}}|t|^{1-\alpha-\beta+\theta_{5}} + \cdots \mbox{\Large $]$} \nonumber \\
  &  & +\: B_{p}^{\mbox{\scriptsize \bf iv}}|t|^{1-\alpha+\gamma}\mbox{\Large $[$}1 + b_{p}^{\mbox{\scriptsize \bf iv}}|t|^{\theta} + \cdots \mbox{\Large $]$},  \label{rhoc-p1}
 \end{eqnarray}
where the leading amplitudes are
 \begin{equation}
  A_{p}^{\mbox{\scriptsize \bf iv}} = QW_{+0}^{0}|\tau|^{2-\alpha}, \hspace{0.1in}  B_{p}^{\mbox{\scriptsize \bf iv}} = Y_{1}|\tau|^{1-\alpha+\gamma}/U,  \label{rhoc-p2}
 \end{equation}
while $\check{p}_{\sigma,1}$, $\check{p}_{\sigma,2}$, $\tau$ and the correction amplitudes are give in (\ref{p-coeff-a}), (\ref{tau1}) and {\bf K}(3.139). The amplitude $A_{p}^{(\mbox{\scriptsize \bf iv})}$ is positive by convexity. 

In the $(\mu,T)$ plane, the critical isochore is given by the closely analogous form
 \begin{eqnarray}
  \check{\mu}^{\mbox{\scriptsize \bf iv}}(t) & = & [\mu^{\mbox{\scriptsize \bf iv}}(T)-\mu_{\mbox{\scriptsize c}}]/k_{\mbox{\scriptsize B}}\Tc  \nonumber \\
   & = & \check{\mu}_{\sigma,1}t + \check{\mu}_{\sigma,2}t^{2} +\cdots + j_{2}A_{p}^{\mbox{\scriptsize \bf iv}}|t|^{2-\alpha}\mbox{\Large $[$} 1 + a_{1p}^{\mbox{\scriptsize \bf iv}}|t|^{\theta} \nonumber \\
  &  & +\: a_{2p}^{\mbox{\scriptsize \bf iv}}|t|^{\gamma-\alpha} + \cdots  + a_{3p}^{\mbox{\scriptsize \bf iv}}|t|^{1-\alpha-\beta+\theta_{5}} + \cdots\mbox{\Large $]$} \nonumber \\
  &  & +\: (1+j_{2})B_{p}^{\mbox{\scriptsize \bf iv}}|t|^{1-\alpha+\gamma}\mbox{\Large $[$}1 + b_{p}^{\mbox{\scriptsize \bf iv}}|t|^{\theta} + \cdots \mbox{\Large $]$},  \label{rhoc-mu1}
 \end{eqnarray}
in which $\check{\mu}_{\sigma,1}$ and $\check{\mu}_{\sigma,2}$ are given in (3.16) and {\bf K}(3.78), while the coefficients $A_{p}^{\mbox{\scriptsize \bf iv}}$, $a_{1p}^{\mbox{\scriptsize \bf iv}}$, etc. are the {\em same} as in (4.28).

When the pressure mixing coefficient $j_{2}$ vanishes, the leading singular $|t|^{2-\alpha}$ term here vanishes; but in that case the third-derivative of $\mu^{\mbox{\scriptsize \bf iv}}(T)$, the chemical potential on the critical isochore above $\Tc$, diverges at criticality, since one typically has $2< 1-\alpha +\gamma < 3$. In view of these results one might also define a magnitude for a Yang-Yang-type of anomaly by comparing the leading singularities in $p$ and $\mu$ on the critical isochore {\em above} $\Tc$; but this naturally leads to the {\em identical} ratio! One may thus write ${\cal R}_{\mu}^{+} = -j_{2}/(1-j_{2}) = {\cal R}_{\mu}$: see (\ref{yang-anomaly2}).

\subsection{{\boldmath $k$}-susceptibility loci}
\label{sec4.5}
In this section we focus on the novel $k$-loci defined in the one-phase region via the isothermal maxima of $\chi^{(k)} \equiv \chi/\rho^{k}$ in the $(\rho,T)$ plane \cite{ork:fis:pan}, where $\chi = (\partial\rho/\partial\mu)_{T}$: see also (\ref{chi}). If one considers $\chi^{(k)}$ as a function of $\mu$ and $T$, the maxima of $\chi^{(k)}$ at fixed $T$ satisfy
 \begin{equation}
  \left(\frac{\partial\chi^{(k)}}{\partial\mu}\right)_{T} = \frac{1}{\rho^{k}}\left[ \left(\frac{\partial\chi}{\partial\mu}\right)_{T} - k\frac{\chi}{\rho}\left(\frac{\partial\rho}{\partial\mu}\right)_{T}\right] = 0.  \label{chi-eq1}
 \end{equation}
This leads to the condition
 \begin{equation}
  \rho\left(\frac{\partial\chi}{\partial\mu}\right)_{T} = k \chi^{2} \hspace{0.1in} \mbox{or} \hspace{0.1in}   \check{\rho}\left(\frac{\partial\check{\chi}_{NN}}{\partial\check{\mu}}\right)_{T} = k (\check{\chi}_{NN})^{2};  \label{chi-eq2}
 \end{equation}
see (\ref{red-den}) and (\ref{suscept}). By using (\ref{chi-gen}), one can obtain the $\check{\mu}$-derivative of $\check{\chi}_{NN}$ and express the $k$-locus in terms of the generalized susceptibilities introduced in (2.28). After some algebra, one finds
 \begin{equation}
   \tilde{\chi}_{hhh} + (1+j_{2})\tilde{\rho}\tilde{\chi}_{hhh} - e_{0}(k)\tilde{\chi}_{hh}^{2} - 3e_{4}\tilde{\chi}_{hht} + \cdots = 0,  \label{chi-cond}
 \end{equation}
where, by convention, $\tilde{\chi}_{hhh} \equiv (\partial^{3}\tilde{p}/\partial\tilde{h}^{3})_{\tilde{t}}$ and $\tilde{\chi}_{hht} \equiv (\partial^{3}\tilde{p}/\partial\tilde{h}^{2}\partial\tilde{t})$, while
 \begin{equation}
  e_{0}(k) = 3j_{2} + k(1-j_{2}) \hspace{0.1in} \mbox{and} \hspace{0.1in} e_{4} = (l_{1}+j_{1})/(1-j_{2}).  \label{chi-cond-a}
 \end{equation}
The condition may now be converted to scaling form by using (2.3) and (2.4); then by employing the small-$y$ expansions (2.8) one can solve to obtain
 \begin{equation}
  y \equiv U\tilde{h}/|\tilde{t}|^{\Delta} = e_{0}(k)\bar{Y}_{1}|\tilde{t}|^{\beta} + \bar{Y}_{2}|\tilde{t}|^{1-\alpha-\beta}
+ \cdots,  \label{chi-y}
 \end{equation}
which might be compared with (\ref{rhoc-eq3}) for the critical isochore. The coefficients are given by
 \begin{equation}
  \bar{Y}_{1} = \mbox{$\frac{1}{6}$} QU(W_{+2}^{0})^{2}/W_{+4}^{0}, \hspace{0.2in} \bar{Y}_{2} = \mbox{$\frac{1}{4}$} \gamma e_{4}W_{+2}^{0}/UW_{+4}^{0},  \label{chi-ycoeff}
 \end{equation}
and we may note, for its future significance, that the leading term in (4.35) vanishes identically for the special $k$ value
 \begin{equation}
  k_{\mbox{\scriptsize opt}} = -3j_{2}/(1-j_{2}) = 3{\cal R}_{\mu}.  \label{chi-ycoeff-b}
 \end{equation}

At this point the scaling field $\tilde{p}$ can be expanded in powers of $y$ and $|\tilde{t}|$ via (2.3) and (2.8) and then, via (4.35), wholly in terms of $|\tilde{t}|$. Finally, by using (\ref{tilde-g})-(\ref{tilde-t}), and rewriting (4.35) as an expansion for $\tilde{h}$ one can solve for $\check{p}$ iteratively as a function of $t$. After some algebra, we find
 \begin{eqnarray}
  \check{p}^{(k)}(t) & = & \check{p}_{\sigma,1}t + \check{p}_{\sigma,2}t^{2} + \cdots + A_{p}^{(k)}|t|^{2-\alpha}\left[ 1 + a_{1p}^{(k)}|t|^{\theta} \right. \nonumber \\
  &  & \hspace{0.1in} \left. +\: a_{2p}^{(k)}|t|^{2\beta} + \cdots + a_{3p}^{(k)}|t|^{\beta+\theta_{5}} + \cdots \right] \nonumber \\
  &  & \hspace{0.1in} +\: B_{p}^{(k)}|t|^{1+\gamma-\alpha} + \cdots,   \label{chi-checkp}
 \end{eqnarray}
where the leading amplitudes are
 \begin{eqnarray}
  A_{p}^{(k)} & = & QW_{+0}^{0}|\tau|^{2-\alpha}\left[ 1 + e_{0}(k)(W_{+2}^{0})^{2}/6W_{+0}^{0}W_{+4}^{0} \right],\nonumber \\ \label{chi-checkp-coeff}  \\
 B_{p}^{(k)} & = & \bar{Y}_{2}|\tau|^{1+\gamma-\alpha}/U,  \label{chi-checkp-coeff-a}
 \end{eqnarray}
while $\check{p}_{\sigma,1}$, $\check{p}_{\sigma,2}$, $\tau$ and the remaining coefficients are given in (3.12), (3.14) and {\bf K}(3.153).

The $k$-loci in the $(\mu,T)$ plane can now be obtained by substituting this result into the scaling fields and using (4.35) once more. The result is
 \begin{eqnarray}
  \check{\mu}^{(k)}(t) & = & \check{\mu}_{\sigma,1}t + \check{\mu}_{\sigma,2}t^{2} + \cdots + A_{\mu}^{(k)}|t|^{2-\alpha}\left[ 1 + a_{1\mu}^{(k)}|t|^{\theta} \right. \nonumber \\
  & & \hspace{0.1in} \left. +\: a_{2\mu}^{(k)}|t|^{2\beta} + \cdots + a_{3\mu}^{(k)}|t|^{\beta+\theta_{5}} + \cdots \right] \nonumber \\
  &  & \hspace{0.1in}+\: B_{\mu}^{(k)} |t|^{1+\gamma-\alpha} + \cdots,  \label{chi-checkmu}
 \end{eqnarray}
where $\check{\mu}_{\sigma,1}$ and $\check{\mu}_{\sigma,2}$ are given in (3.16) and {\bf K}(3.78) while the principal amplitudes are
 \begin{eqnarray}
 A_{\mu}^{(k)} & = & j_{2}A_{p}^{(k)} + e_{0}(k)\bar{Y}_{1}|\tau|^{2-\alpha}/U, \nonumber \\
 B_{\mu}^{(k)} & = & (1+j_{2})B_{p}^{(k)},  \label{chi-checkmu-coeff} 
 \end{eqnarray}
and $\,a_{i\mu}^{(k)} = j_{2}A_{p}^{(k)}a_{ip}^{(k)}/A_{\mu}^{(k)}\,$ for $i=1,\,2,\,3$.

For practical purposes the form of the $k$-loci in the density-temperature plane is of most interest. To that end, note that the generalized densities, $\tilde{\rho}$ and $\tilde{s}$, can be written using the above results as
 \begin{eqnarray}
  \tilde{\rho} & = & 2QU\mbox{\Large $[$} e_{0}(k)\bar{Y}_{1}W_{+2}^{0}|\tilde{t}|^{2\beta} + \bar{Y}_{2}W_{+2}^{0}|\tilde{t}|^{1-\alpha} + \cdots  \nonumber \\
  &  &  \hspace{0.35in} +\: e_{0}(k)\bar{Y}_{1}U_{4\mbox{\scriptsize c}}W_{+2}^{(4)}|\tilde{t}|^{2\beta+\theta} + \cdots \nonumber \\
  &  & \hspace{0.35in} +\: \mbox{$\frac{1}{2}$} U_{5\mbox{\scriptsize c}}W_{+1}^{(5)}|\tilde{t}|^{\beta+\theta_{5}} + \cdots  \mbox{\Large $]$},   \label{chi-tilderho} \\
  \tilde{s} & = & Q|\tilde{t}|^{1-\alpha}\left[ (2-\alpha)W_{+0}^{0} - \gamma [e_{0}(k)\bar{Y}_{1}]^{2} W_{+2}^{0}|\tilde{t}|^{2\beta} + \cdots \right],\nonumber \\  \label{chi-tildes}
 \end{eqnarray}
where $\tilde{t} \approx \tau t$ with $\tau$ defined in (\ref{tau1}). The $k$-loci in the $(\rho,T)$ plane are given by
 \begin{equation}
  \check{\rho}^{(k)}(t) =  1 + (1-j_{2})\tilde{\rho} + j_{2}(j_{2}-1)\tilde{\rho}^{2} - (l_{1}+j_{1})\tilde{s} + \cdots, 
 \end{equation}
so that finally the $k$-locus varies as
 \begin{eqnarray}
 \rho^{(k)}(T) & = & \rhoc \left\{ 1 + B_{d}^{(k)}|t|^{2\beta} \left[ 1 + b_{4}|t|^{\theta} + \cdots \right] + A_{d}^{(k)}|t|^{1-\alpha} \right. \nonumber  \\
  &  & \hspace{0.2in} \left. +\: A_{1}^{(k)}t + \cdots + B_{5}^{(k)}|t|^{\beta+\theta_{5}} + \cdots \right\},  \label{chi-checkrho} \label{chi-checkrho-a}
 \end{eqnarray}
where the coefficients are
 \begin{eqnarray}
  B_{d}^{(k)} & = & 2e_{0}(k)(1-j_{2})\bar{Y}_{1}QUW_{+2}^{0}|\tau|^{2\beta}, \nonumber \\
  b_{4} & = & U_{4\mbox{\scriptsize c}}W_{+2}^{(4)}|\tau|^{\theta}/W_{+2}^{0}, \label{chi-checkrho-coeff-a} \\
  A_{d}^{(k)} & = & -(l_{1}+j_{1})Q|\tau|^{1-\alpha} \nonumber \\
 &  & \times \left[ (2-\alpha)W_{+0}^{0} + \mbox{$\frac{1}{2}$}\gamma (W_{+2}^{0})^{2}/W_{+4}^{0} \right],  \label{chi-checkrho-coeff} \\
 B_{5}^{(k)} & = &  (1-j_{2})QUU_{5\mbox{\scriptsize c}}W_{+1}^{(5)}|\tau|^{\beta+\theta_{5}}, \label{chi-checkrho-coeff-b}
 \end{eqnarray}
while the expression for $A_{1}^{(k)}$ is rather complicated. What is significant is that the leading amplitude, $B_{d}^{(k)}$, varies linearly with $k$ and vanishes identically when $k$ takes the ``optimal value'' $k_{\mbox{\scriptsize opt}} = 3{\cal R}_{\mu}$ given in (4.37), while the other coefficients exhibited do {\em not} vary with $k$ [despite the superscript label $(k)$]. For $k=k_{\mbox{\scriptsize opt}}$ we may say, loosely, that the $k$-locus points most directly to the critical density $\rhoc$. Indeed, for this reason examining the $k$-loci may be of value in analyzing both experimental and simulation data. Furthermore the relation to the Yang-Yang anomaly ratio is again revealing and suggestive.

Orkoulas, Fisher and Panagiotopoulos \cite{ork:fis:pan} examined the $k$-loci for the hard-core square-well fluid using grand canonical Monte Carlo simulations. They observed that the $k$-loci for different system sizes settle down and become independent of size at high enough temperatures. Within the precision attainable these loci can be considered as the true $k$-loci (for the thermodynamic limit). However, when $T\rightarrow \Tc$, the finite-size loci clearly deviate from the limiting behavior. For the data in hand the finite-size effects become evident when $t = (T-\Tc)/\Tc < 0.1$.

\begin{figure}[h]
\vspace{-0.8in}
\centerline{\epsfig{figure=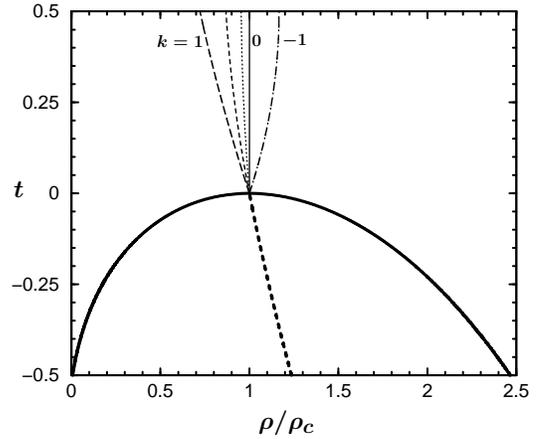,width=3.2in,angle=0}}
\vspace{-0.8in}
 \caption{Selected $k$-susceptibility loci for the van der Waals equation. The thick solid line represents the coexistence curve while the thick dashed line is the diameter. The $(k=0)$ locus is the vertical solid line, while the $k=1,\,\frac{3}{5},\,\frac{1}{4}$ and $k=-1$ loci are portrayed by long dashed, short dashed, dotted, and dot-dashed lines, respectively. Note that the $k=\frac{3}{5}$ locus has the same slope at criticality as the coexistence curve diameter.}
\label{fig3}
\end{figure}

To estimate the $k$-loci near the critical point, we have fitted the data for $t \gtrsim 0.1$ with the formula (4.46)
retaining only the coefficients $B_{d}^{(k)}$, $A_{d}^{(k)}$ and $A_{1}^{(k)}$, while adopting Ising values for the exponents and taking $\rhoc = 0.3067$ and $\Tc = 1.2179$ \cite{ork:fis:pan}. Some of these estimates are presented in Fig.\ \ref{fig1}. Similarly, some of the estimated $k$-susceptibility loci for the restricted primitive model \cite{luijten2} are shown in Fig.\ \ref{fig2}. For comparison, Fig.\ \ref{fig3} presents the $k$-loci for the van der Waals fluid: see Sec.\ \ref{sec5}.
 Note that for the van der Waals fluid all the $k$-loci approach the critical point linearly, which, of course, is consistent with the classical exponent equalities $2\beta = 1-\alpha = 1$.

\subsection{{\boldmath $k$}-heat-capacity loci}
\label{sec4.6}
It is also interesting to examine the $k$-heat-capacity loci or $C_{V}^{(k)}$-loci defined by points of maxima of the modified specific heat
 \begin{equation}
  C_{V}^{(k)}(T,\rho) \equiv C_{V}(T,\rho)/\rho^{k}, \label{ch3-sp1}
 \end{equation}
in the $(\rho,T)$ plane, where $C_{V}$ is the constant-volume specific heat. These are clearly quite analogous to the $k$-susceptibility loci discussed in the previous section; but they have not, as yet, been used in any simulations. The $C_{V}^{(k)}$-loci can, in principle, be obtained in a way similar to that used for the $k$-susceptibility loci by starting with the relation (\ref{red-sp2}). However, when one takes a derivative of (\ref{red-sp2}) with respect to $\mu$ at fixed $T$, the expression becomes complicated and difficult to handle. Therefore we outline a different, canonical approach.

The required maximal points in the $(\rho,T)$ plane at fixed $T$ satisfy 
 \begin{equation}
  -\: kC_{V} + \rho \left(\partial C_{V}/\partial\rho\right)_{T} = 0. \label{cv-eq1}
 \end{equation}
To find a convenient expression for $C_{V}(T,\rho)$, we consider the Helmholtz free energy density $f(\rho,T) = \rho\mu(\rho,T)-p(\rho,T)$. In terms of the reduced variables $\check{\rho}$, $\check{\mu}$ and $\check{p}$ one has
 \begin{eqnarray}
  \check{f}(\check{\rho},t) & \equiv & \left[ f-f_{\mbox{\scriptsize c}}\right]/\rhoc k_{\mbox{\scriptsize B}}\Tc \nonumber \\
  & = & (\mu_{\mbox{\scriptsize c}}/k_{\mbox{\scriptsize B}}\Tc)\Delta\check{\rho} + \check{\rho}\check{\mu} - \check{p} \hspace{0.15in} \mbox{with $~~\Delta\check{\rho} \equiv \check{\rho}-1$.} \nonumber \\ \label{checkf}
 \end{eqnarray}
The reduced specific heat, (\ref{red-sp}), is then
 \begin{equation}
  \check{C}_{V}(\check{\rho},t) = (\rho\Tc/k_{\mbox{\scriptsize B}}\rhoc T)C_{V} = -\left(\partial^{2}\check{f}/\partial t^{2}\right)_{\rho},  \label{check-Cv-free}
 \end{equation}
while the $k$-locus equation, (\ref{cv-eq1}), becomes
 \begin{equation}
  -(k+1)\check{C}_{V} + \check{\rho}\left(\partial \check{C}_{V}/\partial \check{\rho}\right)_{t} = 0.  \label{cv-eq2}
 \end{equation}

To solve this for $\check{\rho}$ as a function of $t$, we first expand $\check{f}$, about the critical density in powers of $\Delta\check{\rho}$ for $t>0$. If the expansion coefficients are $\check{f}_{0}(t)$, $\check{f}_{1}(t)$, $\cdots$, we can rewrite the locus equation as
 \begin{equation}
  -(k+1)\check{f}_{0}^{\prime\prime} + \check{f}_{1}^{\prime\prime} + [(k-1)\check{f}_{1}^{\prime\prime} + 2\check{f}_{2}^{\prime\prime}]\Delta\check{\rho} + {\cal O}\hspace{-0.05in}\left((\Delta\check{\rho})^{2}\right) = 0.  \label{sec3.4.6-eq5}
 \end{equation}

To expand the coefficients $\check{f}_{0}^{\prime\prime}(t)$, etc., in powers of $t$, notice first that, from (4.52) with $\check{\rho}=1$ or $\Delta\check{\rho}=0$, we have
 \begin{equation}
  \check{f}_{0}(t) = \check{\mu}^{\mbox{\scriptsize \bf iv}}(t) - \check{p}^{\mbox{\scriptsize \bf iv}}(t),  \label{sec3.4.6-eq6}
 \end{equation}
where $\check{\mu}^{\mbox{\scriptsize \bf iv}}(t)$ and $\check{p}^{\mbox{\scriptsize \bf iv}}(t)$ represent the variation of $\mu$ and $p$ on the critical isochore as obtained in (\ref{rhoc-p1}) and (\ref{rhoc-mu1}). Then the relation $\mu(T,\rho) = (\partial f/\partial \rho)_{T}$ yields
 \begin{equation}
  \check{f}_{1}(t) = (\mu_{\mbox{\scriptsize c}}/k_{\mbox{\scriptsize B}}\Tc) + \check{\mu}^{\mbox{\scriptsize \bf iv}}(t).  \label{sec3.4.6-eq7}
 \end{equation}
Finally, we have
 \begin{equation}
  \check{f}_{2}(t) = \mbox{$\frac{1}{2}$}\left.\left(\partial\check{\mu}/\partial\check{\rho}\right)_{t}\right|_{\rho=\rho_{\mbox{\tiny c}}} = \mbox{$\frac{1}{2}$} \check{\chi}_{NN}^{-1}(t;\rho=\rhoc).  \label{sec3.4.6-eq8}
 \end{equation}
To obtain the reduced susceptibility, $\check{\chi}_{NN}$, on the critical isochore, we may use (2.29) and the previous results in Sec.\ IV.D. After some algebra we obtain
 \begin{eqnarray}
  \check{\chi}_{NN} & = & 2(1-j_{2})^{2}QU^{2}W_{+2}^{0}|\tau t|^{-\gamma} \nonumber \\
  &  & \times \left[ 1 + U_{4\mbox{\scriptsize c}}(W_{+2}^{(4)}/W_{+2}^{0})|\tau t|^{\theta} + \cdots \right],  \label{check-chiNN2}
 \end{eqnarray}
where $\tau$ was defined in (\ref{tau1}). Note that the pressure-mixing coefficient $j_{2}$ first enters in a $t^{1-\alpha}$ correction; but that is of higher order than the $t^{\theta}$ term retained here. On taking the reciprocal and differentiating twice with respect to $t$, we finally have
 \begin{equation}
  2\check{f}_{2}^{\prime\prime}(t) = D t^{\gamma-2}\left[1 - d_{\theta} t^{\theta} + \cdots \right],  \label{checkmu1}
 \end{equation}
where the coefficients are
 \begin{eqnarray} 
  D & = & \frac{\gamma(\gamma-1)|\tau|^{\gamma}}{2(1-j_{2})^{2}QU^{2}W_{+2}^{0}}, \nonumber \\
  d_{\theta} & = & \frac{(\gamma+\theta)(\gamma+\theta-1)U_{4\mbox{\scriptsize c}}W_{+2}^{(4)}}{\gamma(\gamma-1)W_{+2}^{0}}|\tau|^{\theta}.  \label{checkmu1-coeff}
 \end{eqnarray}

We are now in a position to solve (\ref{sec3.4.6-eq5}) iteratively for $\Delta\check{\rho}$ as a function of $t$ by using (\ref{rhoc-p1}), (\ref{rhoc-mu1}) and (\ref{checkmu1}). One finally obtains the $k$-heat-capacity or $C_{V}^{(k)}$-locus in the form
 \begin{equation}
  \rho_{C}^{(k)} (T)  = \rhoc \left[ 1 - B_{C}^{(k)}t^{2\beta} - A_{C}^{(k)}t^{2\beta+\alpha} + \cdots \right],  \label{cv-check-Drho}
 \end{equation}
where, recalling (4.29), (3.12) and {\bf K}(3.68)-(3.78), the amplitudes are
 \begin{eqnarray}
  B_{C}^{(k)} & = & (2-\alpha)(1-\alpha)[1+k(1-j_{2})]A_{p}^{\mbox{\scriptsize \bf iv}}/D,  \nonumber \\
  A_{C}^{(k)} & = & [\check{p}_{\sigma,2} + k(\check{p}_{\sigma,2}-\check{\mu}_{\sigma,2})]/D.  \label{cv-check-Drho-coeff}
 \end{eqnarray}

Note first that the exponents $2\beta$ and $2\beta + \alpha$ are numerically very close (since $\alpha \simeq 0.1$), so the two terms derived compete strongly near the critical point; furthermore, a $|t|^{1-\alpha}$ term also appears in the full expression. However, the leading amplitude $B_{C}^{(k)}$ does vanish when
 \begin{equation}
  k= k_{C} = -\: 1/(1-j_{2}) = {\cal R}_{\mu} -1.  \label{cv-k0}
 \end{equation}
This value can be compared to $3{\cal R}_{\mu}$ for the $k$-loci: see (4.37).  However, one must keep in mind that it will be more difficult to resolve the `optimal' $k$ value here, compared to the $k$-susceptibility loci, since the term varying as $t^{2\beta+\alpha}$ does not vanish at $k = k_{C}$. 

\begin{figure}[h]
\vspace{-0.8in}
\centerline{\epsfig{figure=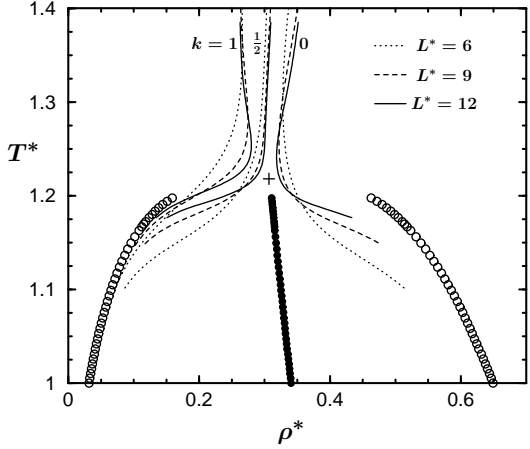,width=3.2in,angle=0}}
\vspace{-0.8in}
 \caption{The $k$-heat-capacity loci for the hard-core square-well fluid in a finite periodic cube of side $L$ derived from previous simulation data [6]. The dotted curves are the loci for $L^{\ast}\equiv L/\sigma = 6$, where $\sigma$ is the diameter of the hard spheres; the dashed lines are for $L^{\ast}=9$, while the solid lines are for $L^{\ast}=12$. Note that since the systems are finite these loci (and the $k$-susceptiblity loci) extend below $\Tc$.}
\label{fig4}
\end{figure}

In order to gain some impression of the behavior of these $C_{V}^{(k)}$-loci, we present in Fig.\ \ref{fig4}, some of the $k$-heat-capacity loci obtained for the hard-core square-well fluid in {\em finite} systems.
 Note that since the exponents $2\beta$, $2\beta +\alpha$, $1-\alpha$, etc., are closely spaced, it is not feasible to extract reliable   estimates of the thermodynamic limiting loci from such finite-size data.

\section{Critical Loci in Classical Theory}
\label{sec5}

In this section, as concrete, albeit rather special, examples of the various critical loci discussed above, we consider the classical theory of a liquid-gas critical point. In particular, the van der Waals equation will be used to provide quantitative illustrations: it reads
 \begin{equation}
  p = \rho k_{\mbox{\scriptsize B}}T/(1-b\rho) - A\rho^{2},  \label{eq5.0.1}
 \end{equation}
where $b$ and $A$ are constants which measure the molecular size and the strength of the attractive interactions, respectively. As well known, one has
 \begin{equation}
  k_{\mbox{\scriptsize B}}\Tc = 8A/27b, \hspace{0.1in} \rhoc = 1/3b, \hspace{0.1in} p_{\mbox{\scriptsize c}}=A/27b^{2}.  \label{eq5.0.1a}
 \end{equation}
Less well known is the behavior when $T\rightarrow 0$ of the liquid and vapor densities and of the phase boundary $\mu_{\sigma}(T) \rightarrow -A/b$: see {\bf K}(App. D).

\subsection{Phase boundaries}
\label{sec5.1}

The Helmholtz free energy (or its appropriate analog) in a classical or Landau theory may be taken as analytic throughout the critical region. Accordingly, to formulate the theory we expand the free energy density around the critical point in terms of the order parameter
 \begin{equation}
  m \equiv (\rho - \rhoc)/\rhoc  \label{eq5.1.1}
 \end{equation}
and the temperature deviation $t=(T-\Tc)/\Tc$, as
 \begin{eqnarray}
  f(T,\rho) & = & \sum_{j=0}^{\infty}\sum_{k=0}^{\infty}a_{jk}m^{j}t^{k}, \nonumber \\
  &  & \mbox{with} \hspace{0.1in} a_{20}=a_{30}=0, \hspace{0.1in} a_{40}>0,  \label{eq5.1.2}
 \end{eqnarray}
where the conditions stated serve merely to ensure normal critical behavior. The explicit values of the leading coefficients $a_{jk}$ for the van der Waals equation (5.1) are derived in {\bf K}(App.\ C) where it is seen that a special feature is that all the {\em cubic coefficients}, $a_{3,k}$ $(k=0,1,2,\cdots)$, {\em vanish identically}, as do many higher order coefficients such as $a_{j2}$, $a_{j3}$, etc., for all $j\geq 2$. Of course, these features should not be expected to hold in real systems or in more realistic models even when a classical description of criticality may be warranted. The leading nonvanishing van der Waals coefficients are reproduced in the Appendix here.

From (\ref{eq5.1.2}), the chemical potential and the pressure can be expanded using
 \begin{equation}
  \mu(T,\rho) = \left(\partial f/\partial\rho\right)_{T}, \hspace{0.1in}  p(T,\rho) = \rho\mu- f.   \label{eq5.1.5}
 \end{equation}
In the two-phase region, the liquid and vapor phases, with densities $\rho_{\mbox{\scriptsize liq}} = \rhoc(1+m_{\mbox{\scriptsize liq}})$ and $\rho_{\mbox{\scriptsize vap}} = \rhoc (1+m_{\mbox{\scriptsize vap}})$, respectively, must have the same chemical potential, $\mu_{\sigma}(T)$, and pressure, $p_{\sigma}(T)$. Solving these two conditions for $\rho_{\mbox{\scriptsize liq}}$ and $\rho_{\mbox{\scriptsize vap}}$ by using (5.3) and (5.4) yields the desired phase boundaries. The detailed calculations are presented in  {\bf K} \cite{yckim}; the results are
 \begin{eqnarray}
   \bar{m}(T) & \equiv & \mbox{$\frac{1}{2}$}(m_{\mbox{\scriptsize liq}}+m_{\mbox{\scriptsize vap}}) = \bar{A}_{1}t + \bar{A}_{2}t^{2} + {\cal O}(t^{3}), \label{eq5.1.8a} \\
   m_{0}(T) & \equiv & \mbox{$\frac{1}{2}$}(m_{\mbox{\scriptsize liq}}-m_{\mbox{\scriptsize vap}}) = B|t|^{1/2}\left[ 1 + C_{1}t + {\cal O}(t^{2}) \right], \nonumber \\ \label{eq5.1.8b} \\
   \mu_{\sigma}(T) & = & \mu_{\mbox{\scriptsize c}} + \mu_{1}t + \mu_{2}t^{2} + \mu_{3}t^{3} + {\cal O}(t^{4}),  \label{eq5.1.8c} \\
   p_{\sigma}(T) & = & p_{\mbox{\scriptsize c}} + p_{1}t + p_{2}t^{2} + p_{3}t^{3} + {\cal O}(t^{4}),  \label{eq5.1.8d}
 \end{eqnarray}
where the amplitudes $\bar{A}_{1}$, $\bar{A}_{2}$, $B$, etc. are expressed in terms of the coefficients $a_{jk}$ in the Appendix. The results agree with those of Sengers and Sengers \cite{sengers} who, however, give them only to  one order lower in $t$.

\subsection{Critical isochore and analytically continued loci}
\label{sec5.2}

In addition to examining {\bf (i)} the {\em critical isochore} above $\Tc$ one may, for critical points describable by classical theory, always consider the {\em analytical continuation} to $T > \Tc$ of various loci otherwise defined only for $T \leq \Tc$. Specifically, we will study: {\bf (ii)} the {\em continued coexistence curve diameter}, $\bar{\rho}(T)$; {\bf (iii)} the {\em continued saturation chemical potential}, $\mu_{\sigma}(T)$; and {\bf (iv)} the {\em continued vapor pressure line}, $p_{\sigma}(T)$. For nonclassical critical behavior, the analysis of the previous section demonstrates that, in general, these last three continuations cannot be uniquely defined since all carry singularities at $\Tc$.

\begin{figure}[h]
\vspace{-0.8in}
\centerline{\epsfig{figure=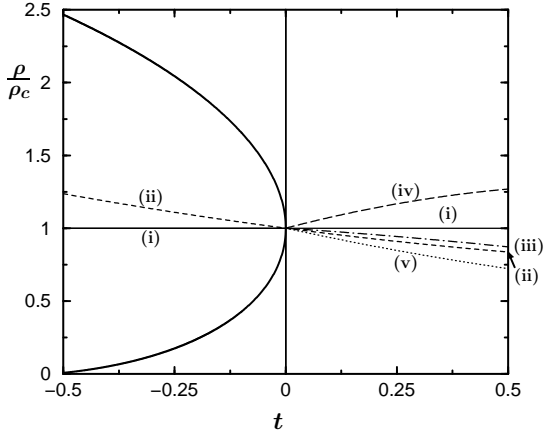,width=3.2in,angle=0}}
\vspace{-0.8in}
\caption{Various critical loci for the van der Waals equation of state in the $(\rho,T)$ plane, where $m=(\rho-\rho_{\mbox{\scriptsize c}})/\rhoc$. The coexistence curve is drawn with a thick solid line. {\bf (i)} Critical isochore; {\bf (ii)} coexistence curve diameter and its analytic continuation into the one-phase region; {\bf (iii)} analytic continuation of $\mu_{\sigma}(T)$; {\bf (iv)} analytic continuation of $p_{\sigma}(T)$; {\bf (v)} $(k=1)$-susceptibility locus.}
\label{fig5}
\end{figure}

We address first the appearance of these loci in the $(\rho,T)$ plane: see Fig.\ \ref{fig5}.
 The critical isochore, {\bf (i)}, is trivial; the continued diameter, say $\rho_{\mbox{\scriptsize \bf ii}}(T)\,\, [\,\equiv \bar{\rho}(T)]$, follows directly from (5.7) and has a critical slope 
 \begin{equation}
   (d\rho_{\mbox{\scriptsize \bf ii}}/dt)_{\mbox{\scriptsize c}} = \bar{A}_{1}\rhoc,
 \end{equation}
which is {\em negative} for the van der Waals equation: see Fig.\ \ref{fig5} for which the quadratic terms in $t$ were also computed.

To determine $\rho_{\mbox{\scriptsize \bf iii}}(T)$, the density locus on which the chemical potential is the analytic continuation of the phase boundary, $\mu_{\sigma}(T)$, we may substitute (\ref{eq5.1.8c}) in the full expansion for $\mu(T,\rho)$ that follows from (5.4) and (5.5). This gives an equation connecting $m$ and $t$ which is easily solved for $m$ in powers of $t$ although the $\mu_{2}t^{2}$ term in (5.8) is needed even in linear order. One finds
 \begin{equation}
  (d\rho_{\mbox{\scriptsize \bf iii}}/dt)_{\mbox{\scriptsize c}} = \rhoc \left(a_{21}a_{50}-2a_{31}a_{40}\right)/8a_{40}^{2}. \label{eq:2.3.4} 
 \end{equation}
The locus $\rho_{\mbox{\scriptsize \bf iii}}(T)$ can be regarded as an {\em effective line of symmetry} along which the chemical potential is analytic. However, as seen in Fig.\ \ref{fig5}, this locus differs from the analytic continuation of the coexistence curve diameter, $\rho_{\mbox{\scriptsize \bf ii}}(T)$ --- another natural candidate --- even in the lowest order in $t$.

The locus $\rho_{\mbox{\scriptsize \bf iv}}(T)$, along which the pressure is the continuation of the vapor pressure curve, $p_{\sigma}(T)$, can be found in an analogous way. Substituting (\ref{eq5.1.8d}) into the expansion of $p$ following from (5.5) leads to
 \begin{eqnarray}
  (d\rho_{\mbox{\scriptsize \bf iv}}/dt)_{\mbox{\scriptsize c}} = \rhoc \left( a_{21}a_{40}+a_{21}a_{50}-2a_{31}a_{40}\right)/8a_{40}^{2}, \label{eq:2.3.6}
 \end{eqnarray}
which, in fact, differs from both (5.10) and (5.11): see Fig.\ \ref{fig5}.

\begin{figure}[h]
\vspace{-0.8in}
\centerline{\epsfig{figure=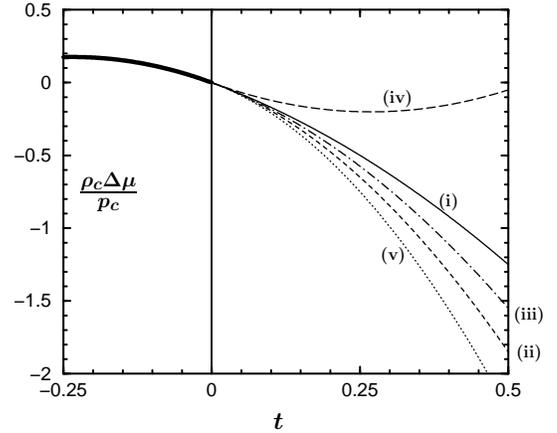,width=3.2in,angle=0}}
\vspace{-0.8in}
\caption{Critical loci for the van der Waals equation in the $(\mu,T)$ plane: $\rhoc\Delta\mu/p_{\mbox{\scriptsize c}}$ is plotted with, for illustrative purposes, $\Delta\mu = \mu - \mu_{\mbox{\scriptsize c}} - (\mu_{1}+\frac{3}{2})t$ [see (A5) and (A13) for $\mu_{\mbox{\scriptsize c}}$ and $\mu_{1}$]. The graphs include only the second and third order terms in $t$. The phase boundary, $\mu_{\sigma}(T)$, is drawn with a thick line. The labeling {\bf (i)}-{\bf (v)} is the same as in Fig.\ \ref{fig5}.}
\label{fig6}
\end{figure}

In the $(\mu,T)$ plane one can find results for the same four loci by using (5.4) and the expressions found for the loci in the $(\rho,T)$ plane. In all cases the initial slopes are the same, that is $(d\mu_{\iota}/dT)_{\mbox{\scriptsize c}} = \mu_{1}$ for $\mbox{$\iota$ = {\bf i} - {\bf iv}}$: see Fig.\ \ref{fig6}.
 However, the initial curvatures differ, being given by $\rhoc\mu_{\mbox{\scriptsize {\bf i},c}}^{\prime\prime} =  2a_{12}$ and
 \begin{eqnarray}
  \rhoc\mu_{\mbox{\scriptsize {\bf ii},c}}^{\prime\prime} & = & 2a_{12}- a_{21}(2a_{31}a_{40}-2a_{21}a_{50})/2a_{40}^{2}, \label{2.29} \\
  \rho_{c}\mu_{\mbox{\scriptsize {\bf iii},c}}^{\prime\prime} & = & 2a_{12}- a_{21}(2a_{31}a_{40}-a_{21}a_{50})/2a_{40}^{2}, \label{2.30}  \\
  \rhoc\mu_{\mbox{\scriptsize {\bf iv},c}}^{\prime\prime} & = & 2a_{12}-a_{21}(2a_{31}a_{40}-a_{21}a_{50}-a_{21}a_{40})/2a_{40}^{2}, \nonumber \\ \label{2.31}
 \end{eqnarray}
where the primes denote differentiation with respect to $t$.

\begin{figure}[h]
\vspace{-0.8in}
\centerline{\epsfig{figure=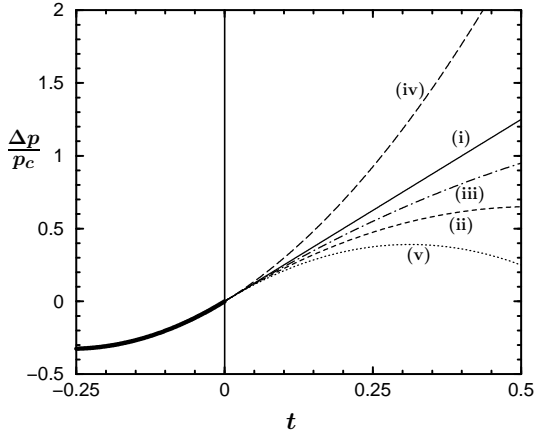,width=3.2in,angle=0}}
\vspace{-0.8in}
\caption{Critical loci for the van der Waals equation in the $(p,T)$ plane, where $\Delta p = p-p_{\mbox{\scriptsize c}}-(p_{1}-2)t$ [see (A8), (A5) and (A12)] and the graphs are correct only up to third order in $t$. The vapor pressure curve, $p_{\sigma}(T)$, is represented by a thick line. The labels {\bf (i)}-{\bf (v)} have the same meaning as in Figs.\ \ref{fig5} and \ref{fig6}.}
\label{fig7}
\end{figure}

Similarly, the loci in the $(p,T)$ plane can be obtained: see Fig.\ \ref{fig7}.
 Again, all have the same initial slopes while the curvatures are distinct as follows from
 \begin{eqnarray}
  p_{\mbox{\scriptsize {\bf i},c}}^{\prime\prime} & = & 2a_{12} -2a_{02}, \label{2.33}  \\
  p_{\mbox{\scriptsize {\bf ii},c}}^{\prime\prime} & = &  2a_{12} -2a_{02} - a_{21}(2a_{31}a_{40}-2a_{21}a_{50})/2a_{40}^{2},  \label{2.34}  \\
  p_{\mbox{\scriptsize {\bf iii},c}}^{\prime\prime} & = &  2a_{12}-2a_{02}-a_{21}(2a_{31}a_{40} - a_{21}a_{50})/2a_{40}^{2},  \label{2.35}  \\
  p_{\mbox{\scriptsize {\bf iv},c}}^{\prime\prime} & = & 2a_{12}-2a_{02}-a_{21}(2a_{31}a_{40} - a_{21}a_{40} \nonumber \\
  &  & \hspace{0.4in} - a_{21}a_{50})/2a_{40}^{2}. \label{2.36}
 \end{eqnarray}

Notice that in Figs.\ \ref{fig6} and \ref{fig7} the critical values as well as conveniently chosen terms linear in $t$ have been subtracted from each locus. Thus one can clearly resolve the differences in curvature and observe that the sequence of loci, from top to bottom, is the same as in Fig.\ \ref{fig5}.

\subsection{{\boldmath $k$}-susceptibility loci}
\label{sec5.2.2}

Since the density increases monotonically with $\mu$ at fixed $T$, the maximal condition (4.31) specifying the $k$-susceptibility loci can be rewritten as
 \begin{equation}
  \left( \frac{\partial\chi^{(k)}}{\partial\rho}\right)_{T} = \frac{1}{\rho^{k}}\left[\left(\frac{\partial \chi}{\partial\rho}\right)_{T} - k \frac{\chi}{\rho} \right] = 0,  \label{ch2-k-loci-eq2}
 \end{equation}
and, thence, in terms of the free energy density $f(T,\rho)$, in the simpler form
 \begin{equation}
  k \left(\partial^{2} f/\partial\rho^{2}\right)_{T} + \rho \left(\partial^{3} f/\partial\rho^{3}\right)_{T} = 0.  \label{ch2-k-loci-eq3}
 \end{equation}
Substituting the expansion (\ref{eq5.1.2}) and solving for $m$ iteratively yields the $k$-loci in the $(\rho,T)$ plane generally as
 \begin{equation}
  \rho^{(k)}(T) = \rhoc\left[ 1 +m_{1}(k)t + m_{2}(k)t^{2} + {\cal O}\hspace{-0.05in}\left( t^{3} \right)\right],  \label{ch2-k-loci-eq5}
 \end{equation}
with coefficients polynomial in $k$, namely,
 \begin{eqnarray}
  m_{1}(k) & = & -(ka_{21} + 3a_{31})/12a_{40},  \label{ch2-k-loci-eq6} \\
  m_{2}(k) & = & -\mbox{$\frac{1}{12}$}\left\{ka_{22}+3a_{32}+3[ (k+1)a_{31}+4a_{41}]m_{1}(k) \right.  \nonumber \\
  &  & \hspace{0.1in} \left. + \: 6[ (k+2)a_{40} + 5a_{50}][ m_{1}(k)]^{2}\right\} /a_{40}.   \label{ch2-k-loci-eq7}
 \end{eqnarray}

For the van der Waals equation these results yield
 \begin{equation}
  m_{1}(k) = -\mbox{$\frac{2}{3}$}k, \hspace{0.1in} m_{2}(k) = -\mbox{$\frac{2}{9}$}k (k^{2} + k -3) \label{ch2-k-loci-eq8}
 \end{equation}
(see Appendix). Note that when $k=0$, the coefficients $m_{1}$ and $m_{2}$ both vanish; furthermore, an exact calculation shows that {\em all} the expansion coefficients vanish identically at $k=0$ so that one has $\rho^{(0)}(T) \equiv \rhoc$: see Fig.\ \ref{fig3}. We also find that the slope of the $k$-locus at criticality becomes equal to that of the coexistence curve diameter when $k=\frac{3}{5}$. In Fig.\ \ref{fig3} the $k$-loci for the van der Waals fluid at several values of $k$ have already been presented. Comparison with Figs.\ \ref{fig1} and \ref{fig2} reveals that the behavior of these loci is closer to those for the hard-core square-well fluid than for the restricted primitive model. Figs.\ \ref{fig5}-\ref{fig7} show how the $(k$$=$$1)$-susceptibility locus appears in the $(\mu,T)$ and $(p,T)$ planes and relates to the other van der Waals loci: see plots labeled {\bf v}.

\subsection{Yang-Yang relation and some extensions}
\label{sec5.3}

It is natural to ask how, if at all, a Yang-Yang anomaly might appear in a classical theory. To that end we discuss, in this section, the Yang-Yang relation (\ref{yang-yang}) on general linear loci in the $(\rho,T)$ plane and also derive an analogous relation for isotherms.

\subsubsection{On the critical isochore}
\label{sec5.3.1}
On the critical isochore $\rho=\rhoc$, in the two-phase region below $\Tc$, it is convenient here to define the functions
 \begin{eqnarray}
  {\cal C}^{-}(T) & \equiv & \rhoc\Tc^{2}\frac{C_{V}}{T}, \hspace{0.1in} {\cal P}^{-}(T) \equiv \Tc^{2}\frac{d^{2}p}{dT^{2}},  \nonumber \\
  & & \hspace{0.1in} {\cal M}^{-}(T) \equiv -\rho_{\mbox{\scriptsize c}}\Tc^{2}\frac{d^{2}\mu}{dT^{2}},   \label{2.42}
 \end{eqnarray}
where the factors $\Tc^{2}$ have been introduced simply for dimensional convenience. [Compare with (3.36)-(3.37).] Above $\Tc$ the functions ${\cal C}^{+}(T)$, ${\cal P}^{+}(T)$ and ${\cal M}^{+}(T)$ may be defined on the critical isochore in precisely the same way. The Yang-Yang relation then implies ${\cal C}^{\pm}={\cal P}^{\pm}+{\cal M}^{\pm}$. In classical theory we may expand in powers of $t$ for $t\gtrless 0$ to obtain
 \begin{equation}
  {\cal C}^{\pm}(T) = {\cal C}_{\mbox{\scriptsize c}}^{\pm} + {\cal C}_{1}^{\pm}t + {\cal O}\left(t^{2}\right),  \label{2.49}
  \end{equation}
and likewise for ${\cal P}^{\pm}(T)$ and ${\cal M}^{\pm}(T)$. The leading amplitudes above $\Tc$ are simply
 \begin{equation}
  {\cal C}_{\mbox{\scriptsize c}}^{+}=-2a_{02}, \hspace{0.1in} {\cal P}_{\mbox{\scriptsize c}}^{+}=2(a_{12}-a_{02}), \hspace{0.1in} {\cal M}_{\mbox{\scriptsize c}}^{+}=-2a_{12}, \label{2.50.3} 
 \end{equation}
while the amplitudes of the terms linear in $t$ are
 \begin{eqnarray}
  {\cal C}_{1}^{-} & = & 6(p_{3}-\rho_{\mbox{\scriptsize c}}\mu_{3}), \hspace{.1in} {\cal C}_{1}^{+} = -6a_{03}, \hspace{0.1in} {\cal P}_{1}^{-} = 6p_{3}, \label{2.51.2} \\
  {\cal P}_{1}^{+} & = & 6(a_{13}-a_{03}),  \hspace{0.1in}  {\cal M}_{1}^{-} =  -6\rho_{\mbox{\scriptsize c}}\mu_{3}, \hspace{.1in} {\cal M}_{1}^{+} = -6a_{13},  \nonumber
 \end{eqnarray}
where $\mu_{3}$ and $p_{3}$ are given in the Appendix from which one readily sees that the slopes are discontinuous across $\Tc$.

Indeed, it is well known that the specific heat $C_{V}(T)$ on the critical isochore exhibits a finite jump at a classical critical point which, in fact, is of magnitude given by
 \begin{equation}
  \rhoc\Tc\Delta C_{V} = \Delta {\cal C} \equiv {\cal C}_{\mbox{\scriptsize c}}^{-} - {\cal C}_{\mbox{\scriptsize c}}^{+} = \mbox{$\frac{1}{2}$}a_{21}^{2}/a_{40}.    \label{2.54}
 \end{equation}
It transpires, however, that in general both the second derivative of the pressure {\em and} of the chemical potential have a finite discontinuity on the critical isochore. This arises from the presence of the odd coefficients $a_{31}$ and $a_{50}$ (even though $a_{31}$ vanishes ``by accident'' for the van der Waals equation) as follows from
 \begin{eqnarray}
  \Delta {\cal M} & \equiv & {\cal M}_{\mbox{\scriptsize c}}^{-} - {\cal M}_{\mbox{\scriptsize c}}^{+} = \mbox{$\frac{1}{2}$}a_{21}(2a_{31}a_{40} - a_{21}a_{50})/a_{40}^{2},  \label{2.53} 
 \end{eqnarray}
with, of course, $\Delta{\cal P} = \Delta {\cal C}-\Delta {\cal M}$.

\begin{figure}[h]
\vspace{-0.8in}
\centerline{\epsfig{figure=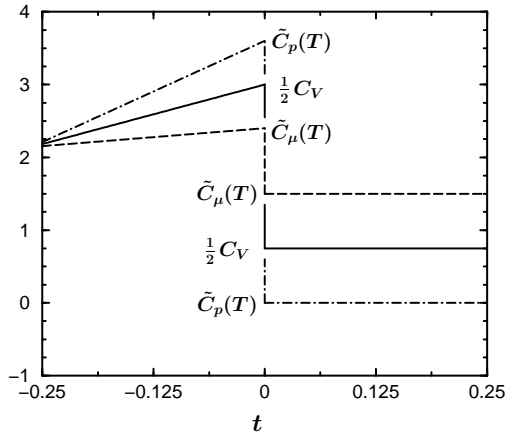,width=3.2in,angle=0}}
\vspace{-0.8in}
\caption{ The specific heat $C_{V}(T,\rhoc)$ (solid curve) on the critical isochore of a van der Waals fluid (times $\frac{1}{2}$ in units of $k_{\mbox{\scriptsize B}}$), compared with the corresponding contributions, $\tilde{C}_{p}(T)$ and $\tilde{C}_{\mu}(T)$ (dot-dashed and dashed plots), due to the isochoric variation of pressure and chemical potential: see (3.37) and (5.26). Note that the standard kinetic contribution to the total specific heat, namely $\frac{1}{2}dk_{\mbox{\scriptsize B}}$ (in $d$ dimensions), arises {\em entirely} from $\tilde{C}_{\mu} = -T\mu^{\prime\prime}$; the plots beneath $\Tc$ are correct only to leading order in $t$.}
\label{fig8}
\end{figure}

These results are illustrated for a van der Waals fluid in Fig.\ \ref{fig8}.
 Clearly, the variation of both the pressure and the chemical potential contribute to the discontinuity in the specific heat on the critical isochore. The relative degree of the two contributions may be gauged by evaluating the ratio
 \begin{equation}
  \dot{\cal R}_{\mu} \equiv \frac{\Delta {\cal M}}{\Delta {\cal C}} = 2\frac{a_{31}}{a_{21}} - \frac{a_{50}}{a_{40}},  \label{2.56}
 \end{equation}
which, as the notation suggests, might be regarded as an effective Yang-Yang ratio for classical systems. Indeed, if $\dot{\cal R}_{\mu}$ vanishes, as in the case of simple lattice gas mean-field models, the isochoric second temperature derivative of the chemical potential becomes continuous through the critical point: then, the only contribution to the discontinuity in the specific heat on the critical isochore arises from the variation of the {\em pressure}.

\subsubsection{On linear density loci}
\label{sec5.3.2}

In light of the various critical loci that approach the critical point linearly in the $(\rho,T)$ plane of classical systems (see Fig.\ \ref{fig5}), it is rather natural to extend the Yang-Yang relation to the general linear locus
 \begin{equation}
  \rho=\rhoc (1+rt). \label{2.57.0}
 \end{equation}
To that end consider, first, the second temperature derivatives of the pressure and the chemical potential near criticality along this locus. For $T<\Tc$, there is no dependence on $r$ because, since the exponent $\beta$ is less than unity, the linear locus (5.33) always lies in the two-phase region when $T\rightarrow\Tc-$. Hence the results (5.29)-(5.31) still apply. In the single-phase region above $\Tc$, the previous definitions may naturally be extended by taking
 \begin{equation}
  {\cal P}^{\pm}_{r}(T) \equiv \Tc^{2} \left(\frac{\partial^{2} p}{\partial T^{2}}\right)_{r},  \hspace{.1in} {\cal M}^{\pm}_{r}(T) \equiv -\rho \Tc^{2}\left(\frac{\partial^{2}\mu}{\partial T^{2}}\right)_{r}. \label{2.58}
 \end{equation}
From (5.4) and (5.5) these functions may be expanded straightforwardly to yield the limiting values 
 \begin{eqnarray}
  {\cal P}_{r,\mbox{\scriptsize c}}^{+} & = & 2(2a_{12}r +a_{12} - a_{02}),\nonumber \\
  {\cal M}_{r,\mbox{\scriptsize c}}^{+} & = & -2(a_{12} + 2a_{21}r). \label{2.62}
 \end{eqnarray}
In an obvious notation, we then obtain
 \begin{eqnarray}
 \Delta{\cal P}_{r} & = &a_{21}(a_{21}a_{40} - 2a_{31}a_{40} + a_{21}a_{50})/2a_{40}^{2} - 4a_{21}r, \nonumber \\ \label{2.65} \\
  \Delta{\cal M}_{r} & = & 4a_{21}r + a_{21}(2a_{31}a_{40} - a_{21}a_{50})/2a_{40}^{2}. \label{2.66}
 \end{eqnarray}
When $r=0$, i.e. along the critical isochore, these results of course agree with (5.31).

It is now interesting to define $r_{\cal P}$ as specifying that locus along which the pressure has a continuous second derivative at criticality and similarly for $r_{\cal M}$. From (\ref{2.65}) and (\ref{2.66}), we thus find
 \begin{eqnarray}
  r_{\cal M} & = & (a_{21}a_{50}-2a_{31}a_{40})/8a_{40}^{2} = r_{\cal P} - \mbox{$\frac{1}{8}$}a_{21}/a_{40}.  \label{2.68}
 \end{eqnarray}
For the van der Waals equation these expressions yield $r_{\cal P}=0.8$ and $r_{\cal M}=-0.2$. On the other hand, from (5.10), the locus of the analytically continued coexistence diameter is specified by
 \begin{equation}
  \bar{r} = (a_{21}a_{50}-a_{31}a_{40})/4a_{40}^{2}, \label{2.69}
 \end{equation}
which takes the value $\bar{r}=-0.4$ for the van der Waals equation. Evidently, all three loci are distinct! Note especially, however, that $r_{\cal M}$ is equal to the slope of the effective line of symmetry at the critical point: see (5.11). Hence, as already shown by Mulholland \cite{mulholland}, this putative line of symmetry for the van der Waals equation differs from the analytic continuation of the coexistence curve diameter; but it is, rather, the locus on which the effective Yang-Yang ratio, $\dot{\cal R}_{\mu}$, vanishes!

Now, extending the Yang-Yang relation to the general locus (5.33) leads to
 \begin{eqnarray}
  \frac{\rho C_{V}}{T} + \frac{\rho_{\mbox{\scriptsize c}}^{2}r^{2}}{T_{\mbox{\scriptsize c}}^{2}\rho^{2}K_{T}} & = & \Tc^{-2}\left[ {\cal P}_{r}^{\pm}(T) + {\cal M}_{r}^{\pm}(T) \right],  \label{2.70}
 \end{eqnarray}
where $\rho=\rho_{\mbox{\scriptsize c}}(1+rt)$ is understood and, as before, $K_{T}$ denotes the isothermal compressibility: see {\bf K}(App. F).

In classical theory, this relationship may be verified straightforwardly using (5.4), and subsequently derived expressions for $\rho C_{V}/T$ and $1/\rho^{2}K_{T}$, on (5.33). Explicitly, one finds the initial derivatives of ${\cal P}_{r}^{+}(T)$ and ${\cal M}_{r}^{+}(T)$ at criticality to be cubic polynomials, namely,
 \begin{eqnarray}
  {\cal P}_{r,1}^{+} & = & 6\left[a_{13}-a_{03} + 2a_{22}r + (a_{21} + 3a_{31})r^{2} + 4a_{40}r^{3}\right], \nonumber \\  \label{2.76} \\
  {\cal M}_{r,1}^{+} & = & -\left[6a_{131} + 2(a_{12} + 6a_{22})r + (4a_{21} + 18a_{31})r^{2} \right. \nonumber \\
  &  & \hspace{0.15in} +\: \left. 24a_{40}r^{3}\right],  \label{2.77}
 \end{eqnarray}
which, of course, satisfy the generalized relation (5.40) with
 \begin{equation}
  \rhoc^{2}/\rho^{2}K_{T}=2a_{21}t + {\cal O}(t^{2}).
 \end{equation}

\subsubsection{Yang-Yang-type relation on the critical isotherm}
\label{sec5.3.4}
An important locus in the $(\rho,T)$ plane is the critical isotherm, $T=\Tc$. In analogy to the Yang-Yang relation for isochoric variations one can derive a corresponding {\em isothermal} relation, namely,
 \begin{equation}
  \frac{1}{\rho^{2} K_{T}} = \left(\frac{\partial^{2} p}{\partial\rho^{2}}\right)_{T} - \rho\left(\frac{\partial^{2}\mu}{\partial\rho^{2}}\right)_{T}:   \label{2.78}
 \end{equation}
see {\bf K}(2.97).

To apply this to the critical isotherm, recall that the generally expected critical behavior is
 \begin{equation}
  1/\rhoc^{2}K_{T} \approx D^{\pm}\left|\Delta\rho\right|^{\delta -1}, \hspace{.1in} \mbox{for $~~\Delta\rho \equiv \rho-\rhoc \rightarrow 0\pm$.}  \label{2.80}
 \end{equation}
In classical theory one has $\delta=3$ and finds $D^{+} = D^{-} = 12a_{40}/\rhoc^{4}.$ More generally one might define the pressure and chemical potential contributions in (5.44) via
 \begin{eqnarray}
  {\cal P}_{\rho}(\rho) & = & \left(\partial^{2} p/\partial\rho^{2}\right)_{T}, \hspace{0.1in} {\cal M}_{\rho}(\rho) = -\left(\partial^{2}\mu/\partial\rho^{2}\right)_{T}, \nonumber \\
  &  & \hspace{0.5in} \mbox{at $\,\, T=\Tc$.}  \label{2.83}
 \end{eqnarray}

It is then natural to ask, as on the critical isochore, how the two contributions --- one from $p$, and one from $\mu$ --- contribute to the overall singularity in (5.45) and whether that might throw any further light on the pressure-mixing coefficients $j_{1}$ or $j_{2}$. It turns out, however, that matters are very different! In fact, by appropriate integration of (\ref{2.80}), one can obtain the leading singular behavior of {\em both} $p$ and $\mu$ on the critical isotherm. Then one finds that the leading singularities of ${\cal P}_{\rho}(\rho)$ and ${\cal M}_{\rho}(\rho)$ both vary as $|\Delta\rho|^{\delta-2}$ and are, thus {\em more singular} than is their sum! Indeed, these leading terms must cancel exactly in (\ref{2.78}) so that the behavior (\ref{2.80}) for $1/K_{T}$ appears only as a net correction term. Needless to say, the classical theory fits in with this description although, since $\delta$ is an odd integer, there is ambiguity in defining appropriate ``singular contributions''. More generally, while there may well be effective ways in which an estimate of the degree of pressure mixing can be found from isothermal observations, we have not identified a good candidate.

\section{Summary}
\label{sec6}
We have carefully formulated a full or ``complete scaling theory'' for asymmetric fluid criticality that, in particular, incorporates pressure mixing in the basic linear and nonlinear scaling fields. The theory can then describe a Yang-Yang anomaly \cite{fis:ork} in which the second temperature derivative of the chemical potential along the phase boundary, namely, $d^{2}\mu_{\sigma}/dT^{2}$, {\em diverges} at the critical point with the specific heat exponent $\alpha$. This is shown to entail, also, a leading $|t|^{2\beta}$ term in the coexistence curve diameter that dominates over the previously known $|t|^{1-\alpha}$ term \cite{mer}. The strength of the Yang-Yang anomaly, ${\cal R}_{\mu}$, is directly related to the linear pressure-mixing coefficient in the ordering field $\tilde{h}$ that we have labeled $j_{2}$. 

For convenience of reference, we identify here the main definitions introduced and the explicit results derived. The coefficients $j_{1}$, $j_{2}$, $k_{0}$, $\cdots$, $n_{5}$, $v_{1}$, $v_{2}$ entering the {\em nonlinear scaling fields} to quadratic order are defined via (1.4)-(1.7), in terms of reduced pressure, $\check{p}$, chemical potential, $\check{\mu}$ and reduced temperature deviation $t=(T-\Tc)/\Tc$: see (1.7). The {\em scaling ansatz}, in the form developed is presented in (2.1)-(2.3) and the basic properties of the overall scaling function $\,W_{\pm}(y,y_{4},y_{5},\cdots)\,$ are set out in (2.4) together with (2.8) and (2.10) for small argument, $|y|\rightarrow 0$, and in (2.12) for $|y|\rightarrow\infty$. Explicit expansions in powers of $t$ for the phase boundaries, $p_{\sigma}(T)$ and $\mu_{\sigma}(T)$, are reported in (3.11) and (3.15). The corresponding expressions for the {\em coexistence curve} (and its diameter), and for the entropies of coexisting gas and liquid are (3.20) and (3.21); Eq.\ (3.31) gives the entropy on the critical isochore, $\rho=\rhoc$. The linear mixing coefficients, themselves are related to measurable critical amplitudes, etc. in (3.22) and (3.32)-(3.35).

Section~\ref{sec4} addresses the behavior of various critical loci in the $\mu$, $p$, and $\rho$ vs.\ $T$ planes. The {\em critical isokyme}, defined by $\mu=\mu_{\mbox{\scriptsize c}}$, is described in (4.5) and (4.7); the {\em critical isobar}, in (4.13) and (4.15); the {\em isotherm}, in (4.19) and (4.21); and the {\em critical isochore} in (4.28) and (4.30). The family of $k$-{\em susceptibility loci} is defined via (4.31) while (4.38), (4.41) and (4.46), with (4.34) elucidate their behavior in the $p$, $\mu$, and density vs.\ $T$ planes, respectively: see Figs.\ \ref{fig1}-\ref{fig3}. The analogous $k$-heat-capacity loci are defined in (4.50) while (4.62) describes their appearance in the $(\rho,T)$ plane: see Fig.\ \ref{fig4}. For both the $k$-susceptibility and $k$-heat-capacity loci there are ``optimal'' values of $k$, depending only on the pressure-mixing coefficient $j_{2}$, for which the dominant singularity at criticality vanishes. Thus these particular loci approach the critical point in the $(\rho,T)$ plane almost linearly: see (4.37) and (4.64).

The magnitudes and inter-relations of the various critical loci, including {\em analytically continued loci}, in the $(\rho,T)$, $(\mu,T)$ and $(p,T)$ planes are illustrated in Figs.\ \ref{fig5}, \ref{fig6} and \ref{fig7}, respectively. Corresponding analytic results for a general classical fluid, defined via the Landau expansion (5.4), are given in (5.6)-(5.9) and (5.10)-(5.19) and, for the $k$-loci, in (5.22). The Yang-Yang relation and its generalization to a linear density locus [in (5.34) and (5.40)] are discussed for classical systems --- for which there is no uniquely defined ``anomaly'' --- in Section V.D. Fig.\ \ref{fig8} illustrates that both pressure and chemical potential variation contribute to the specific-heat jump in a van der Waals fluid.

Finally, in part II of this article \cite{yckim:fisher}, the present scaling formulation with pressure mixing will be extended to {\em finite-size systems} (with periodic boundary conditions). The results, which include the definition and analysis of ``Q-loci'', are of practical importance in the analysis of simulation data for near-critical asymmetric systems \cite{ork:fis:pan,luijten2} as will be further demonstrated \cite{yckim:fisher}.

\acknowledgements

Erik Luijten collaborated in producing the coexistence curve estimates for the restricted primitive model electrolyte presented in Fig.\ \ref{fig2}. Support from the National Science Foundation (through Grant No.\ CHE 99-81772) is gratefully acknowledged.

\appendix
\section*{$~~~$ Phase Boundaries in Classical Theory}
\setcounter{equation}{0}
\renewcommand{\theequation}{A\arabic{equation}}

The amplitudes for the coexistence curve introduced in (5.6) and (5.7) are found to be
 \begin{eqnarray}
  \bar{A}_{1} & = & \left(a_{21}a_{50}-a_{31}a_{40}\right)/4a_{40}^{2}, \label{eq:bb.1} \\
  \bar{A}_{2} & = & \left[a_{31}(4a_{40}^{3}a_{41}-4a_{31}a_{40}^{2}a_{50}+10a_{21}a_{40}a_{50}^{2}\right. \nonumber \\
        &   & -\:6a_{21}a_{40}^{2}a_{60})-4a_{32}a_{40}^{4}+a_{50}(4a_{22}a_{40}^{3} \nonumber \\
  &  & -\: 8a_{21}a_{40}^{2}a_{41} - 6a_{21}^{2}a_{50}^{2}+9a_{21}^{2}a_{40}a_{60}) \nonumber \\
  &  & \left. +\: 4a_{21}a_{40}^{3}a_{51}-3a_{21}^{2}a_{40}^{2}a_{70}\right]/16a_{40}^{5},  \label{eq:bb.2} \\
  B & = & \left[ a_{21}/2a_{40}\right]^{1/2}, \label{eq:bb.3} \\
  C_{1} & = & \left(8a_{22}a_{40}^{3}-8a_{21}a_{40}^{2}a_{41}+6a_{21}^{2}a_{40}a_{60}\right. \nonumber \\
        &   & \left.-\:3a_{31}^{2}a_{40}^{2}+10a_{21}a_{31}a_{40}a_{50}-7a_{21}^{2}a_{50}^{2}\right)/16a_{21}a_{40}^{3}. \nonumber \\ \label{eq:bb.4} 
 \end{eqnarray}
The coefficients introduced in (5.8) and (5.9) for the phase boundaries, $\mu_{\sigma}(T)$ and $p_{\sigma}(T)$, turn out to be
 \begin{eqnarray}
  \rhoc\mu_{\mbox{\scriptsize c}} & = & a_{10},\hspace{.2in} \rhoc\mu_{1} = a_{11}, \label{eq:bb.5} \\
  \rhoc\mu_{2} & = & a_{12}+a_{21}\left(a_{21}a_{50}-2a_{31}a_{40}\right)/4a_{40}^{2},  \label{eq:bb.6} \\
  \rhoc\mu_{3} & = & a_{13}+\left[a_{31}(a_{31}^{2}a_{40}^{3}-4a_{21}a_{31}a_{40}^{2}a_{50}\right. \nonumber \\
                  &   & -\: 4a_{40}^{4}a_{22}+4a_{21}a_{40}^{3}a_{41}+5a_{21}^{2}a_{40}a_{50}^{2} \nonumber \\
  &  & -\: 3a_{21}^{2}a_{40}^{2}a_{60}) - 4a_{32}a_{21}a_{40}^{4}+a_{50}(4a_{21}^{2}a_{40}^{2}a_{41} \nonumber \\
                  &   & -\: 4a_{21}a_{22}a_{40}^{3} + 2a_{21}^{3}a_{50}^{2}-3a_{21}^{3}a_{40}a_{60}) \nonumber \\
  &  & -\: 2a_{51}a_{21}^{2}a_{40}^{3} \left.+a_{70}a_{21}^{3}a_{40}^{2}\right]/8a_{40}^{5}. \label{eq:bb.7}
 \end{eqnarray}
 \begin{eqnarray} 
  p_{\mbox{\scriptsize c}} & = & \rhoc\mu_{0}-a_{00}, \hspace{.2in} p_{1} = \rhoc\mu_{1}-a_{01}, \label{eq:bb.8} \\
  p_{2} & = & \rhoc\mu_{2}-a_{02}+a_{21}^{2}/4a_{40}, \label{eq:bb.9} \\
  p_{3} & = & \rhoc\mu_{3}-a_{03}+a_{21}(4a_{22}a_{40}^{3}-2a_{21}a_{40}^{2}a_{41} \nonumber \\
        &   & +\: a_{21}^{2}a_{40}a_{60}-a_{31}^{2}a_{40}^{2}+2a_{21}a_{31}a_{40}a_{50} \nonumber \\
  &  & -\: a_{21}^{2}a_{50}^{2})/8a_{40}^{4}.  \label{eq:bb.10}
 \end{eqnarray}

To specify the corresponding coefficients $a_{ij}$ for the van der Waals equation [see {\bf K}(App. C)] we take $\Lambda_{\mbox{\scriptsize c}}$ to be the thermal de Broglie wavelength at $T=\Tc$ in a $d$-dimensional system and define
 \begin{equation}
   \lambda = \ln (\Lambda_{\mbox{\scriptsize c}}^{d}/2b) -1.
 \end{equation}
The leading nonvanishing coefficients are then
 \begin{eqnarray}
  a_{00} & = & (\mbox{$\frac{8}{3}$}\lambda -3)p_{\mbox{\scriptsize c}}, \hspace{0.05in} a_{01}=a_{00}-p_{\mbox{\scriptsize c}}, \hspace{0.05in} -a_{02} = 3 a_{03} = 2p_{\mbox{\scriptsize c}}, \nonumber \\  \\
  a_{10} & = & a_{00} + p_{\mbox{\scriptsize c}}, \hspace{0.05in} a_{11} = \mbox{$\frac{8}{3}$} \lambda p_{\mbox{\scriptsize c}}, \hspace{0.05in} -a_{12}=3a_{13} = 2p_{\mbox{\scriptsize c}}, \nonumber \\ \\
  a_{20} & = & 0, \hspace{0.15in} a_{21}=3p_{\mbox{\scriptsize c}}, \hspace{0.15in} a_{40}=a_{41}=\mbox{$\frac{3}{8}$}p_{\mbox{\scriptsize c}}, \\
  a_{50} & = & -\mbox{$\frac{3}{40}$}p_{\mbox{\scriptsize c}}, \hspace{0.15in} a_{60} = \mbox{$\frac{9}{80}$}p_{\mbox{\scriptsize c}}, \hspace{0.15in} a_{70}=-\mbox{$\frac{3}{56}$}p_{\mbox{\scriptsize c}}.
 \end{eqnarray}


\begin{thebibliography}{99}
 \bibitem{yang:yang} C.\ N.\ Yang and C.\ P.\ Yang, Phys.\ Rev.\ Lett.\ {\bf 13}, 303 (1964).
 \bibitem{fis:ork} M.\ E.\ Fisher and G.\ Orkoulas, Phys.\ Rev.\ Lett.\ {\bf 85}, 696 (2000); G.\ Orkoulas, M.\ E.\ Fisher and C.\ \"{U}st\"{u}n, J.\ Chem.\ Phys.\ {\bf 113}, 7530 (2000).
 \bibitem{voronel} M.\ I.\ Bagatskii, A.\ V.\ Voronel' and V.\ G.\ Gusak, Zh.\ Eksp.\ Teor.\ Fiz.\ {\bf 43}, 728 (1962) [Sov.\ Phys. JETP {\bf 16}, 517 (1963)]; A.\ V.\ Voronel' {\em et al.} {\em ibid.} {\bf 45}, 823 (1963) [Sov.\ Phys.\ JETP {\bf 18}, 568 (1964)].
 \bibitem{lee:yan} T.\ D.\ Lee and C.\ N.\ Yang, Phys.\ Rev.\ {\bf 87}, 410 (1952).
 \bibitem{kos:ani} A.\ Kostrowicka Wyczalkowska, M.\ A.\ Anisimov, J.\ V.\ Sengers and Y.\ C.\ Kim, J.\ Chem.\ Phys.\ {\bf 116}, 4202 (2002).
 \bibitem{ork:fis:pan} G.\ Orkoulas, M.\ E.\ Fisher and A.\ Z.\ Panagiotopoulos, Phys.\ Rev.\ E {\bf 63}, 051507 (2001).
 \bibitem{widom} B.\ Widom, J.\ Chem.\ Phys.\ {\bf 43}, 3898 (1965).
 \bibitem{fisher3} M.\ E.\ Fisher, Rep.\ Prog.\ Phys.\ {\bf 30}, 615 (1967).
 \bibitem{kadanoff} L.\ P.\ Kadanoff {\em et al.} Rev.\ Mod.\ Phys.\ {\bf 39}, 395 (1967).
 \bibitem{griffiths} R.\ B.\ Griffiths, Phys.\ Rev.\ {\bf 158}, 176 (1967).
 \bibitem{patashinskii} A.\ Z.\ Patashinskii and V.\ L.\ Pokrovskii, Sov.\ Phys.\ JEPT {\bf 23}, 292 (1966).
 \bibitem{widom2} B.\ Widom, Physica {\bf 73}, 107 (1974).
 \bibitem{sengers} J.\ V.\ Sengers and J.\ M.\ H.\ Levelt Sengers, {\em Progress in Liquid Physics}, edited by C.\ A.\ Croxton (Wiley, Chickester, U.K., 1978) Chap.\ 4.
 \bibitem{sengers2} M.\ S.\ Green, M.\ Vicentini-Missoni and J.\ M.\ H.\ Levelt Sengers, Phys.\ Rev.\ Lett.\ {\bf 18}, 1113 (1967); M.\ Vicentini-Missoni, J.\ M.\ H.\ Levelt Sengers and M.\ S.\ Green, Phys.\ Rev.\ Lett.\ {\bf 22}, 389 (1969); M.\ Vicentini-Missoni, R.\ I.\ Joseph, M.\ S.\ Green and J.\ M.\ H.\ Levelt Sengers, Phys.\ Rev.\ B {\bf 1}, 2312 (1970).
 \bibitem{mer} See, especially, (a) N.\ D.\ Mermin, Phys.\ Rev.\ Lett.\ {\bf 26}, 169 (1971) and (b) J.\ J.\ Rehr and N.\ D.\ Mermin, Phys.\ Rev.\ A {\bf 8}, 472 (1973).
 \bibitem{bruce:wilding} A.\ D.\ Bruce and N.\ B.\ Wilding, Phys.\ Rev.\ Lett.\ {\bf 68}, 193 (1992); N.\ B.\ Wilding and A.\ D.\ Bruce, J.\ Phys.: Condens.\ Matter {\bf 4}, 3087 (1992).
 \bibitem{kim:fis:bar} Y.\ C.\ Kim, M.\ E.\ Fisher and M.\ C.\ Barbosa, J.\ Chem.\ Phys.\ {\bf 115}, 933 (2001).
 \bibitem{fis:bar} M.\ E.\ Fisher and M.\ C.\ Barbosa, Phys.\ Rev.\ B {\bf 43}, 11177 (1991).
 \bibitem{wegner} F.\ J.\ Wegner, Phys.\ Rev.\ Lett.\ {\bf 6}, 1891 (1972); in {\em Phase Transitions and Critical Phenomena}, Edited by C.\ Domb and M.\ S.\ Green (Academic, London, 1976) vol. 6, p. 8.
 \bibitem{fisher1} M.\ E.\ Fisher, ``Scaling, Universality and Renormalization Group Theory,'' in Lecture Notes in Physics, Vol.\ 186, {\em Critical Phenomena}, edited by F.\ J.\ W.\ Hahne (Springer, Berlin, 1983), p.\ 1-139; Rev.\ Mod.\ Phys.\ {\bf 46}, 597 (1974); {\em ibid} {\bf 70}, 653 (1998).
 \bibitem{fisher2} M.\ N.\ Barber, Phys. Repts. {\bf C29}, 1 (1977).
 \bibitem{binney} D.\ I.\ Uzunov, {\em Introduction to the Theory of Critical Phenomena} (World Scientific, Singapore, 1993).
 \bibitem{fis:fel} M.\ E.\ Fisher and B.\ U.\ Felderhof, Ann.\ Phys.\ (N.Y.) {\bf 58}, 176 (1970); {\em ibid} {\bf 58}, 217 (1970).
 \bibitem{fis:kim} M.\ E.\ Fisher and Y.\ C.\ Kim, J.\ Chem.\ Phys.\ {\bf 117}, 779 (2002).
 \bibitem{luijten2} E.\ Luijten, M.\ E.\ Fisher and A.\ Z.\ Panagiotopoulos, Phys.\ Rev.\ Lett.\ {\bf 88}, 185701 (2002).
 \bibitem{yckim} Y.\ C.\ Kim, Ph.\ D.\ Thesis, ``Fluid Criticality: Experiment, Scaling and Simulations,'' University of Maryland (2002). This work, which contains further details of the analyses presented here, will be denoted {\bf K} and equations therein will be referenced as, e.g., {\bf K}(3.41), etc.
 \bibitem{ley-koo} M.\ Ley-Koo and M.\ S.\ Green, Phys.\ Rev.\ A {\bf 23}, 2650 (1981).
 \bibitem{mulholland} G.\ W.\ Mulholland, J.\ Chem.\ Phys.\ {\bf 59}, 2738 (1973).
 \bibitem{yckim:fisher} Y.\ C.\ Kim and M.\ E.\ Fisher, to be published.
\end{thebibliography}
\end{document}